\documentclass[preprint]{elsarticle}
\usepackage{natbib}
\usepackage{amsmath}
\usepackage{indentfirst}
\usepackage{amsfonts}
\usepackage{amssymb}
\usepackage{changepage}
\usepackage{lscape}
\usepackage[]{algorithm2e}
\usepackage{afterpage}
\usepackage{ulem}
\usepackage[T1]{fontenc}
\usepackage{booktabs}

\usepackage{subfigure}
\usepackage{color}
\usepackage{epsfig}
\usepackage{xcolor}
\usepackage{epstopdf}
\usepackage{amsmath}
\usepackage{graphicx,psfrag,epsf}
\usepackage{color}
\usepackage{soul}
\usepackage{multirow}
\usepackage{rotating}
\usepackage{scrextend}
\usepackage{verbatim} 
\usepackage{todonotes}
\usepackage[most]{tcolorbox}
\usepackage{url}

\newcommand{\bE}{\mathbf{E}}

\newcommand{\bY}{\mathbf{Y}}

\newcommand{\Expect}{{\rm I\kern-.3em E}}

\newdefinition{defin}{Definition}

\begin{document}
\begin{frontmatter}
\title{{\textbf{From design of experiments to analysis of variance of multivariate data: a tutorial review on ANOVA simultaneous component analysis.}}}

\author{Jos{\'e} Camacho\corref{cor1}\fnref{fn1}\fnref{fn1b}}
\ead{josecamacho@ugr.es} 
\address{University of Granada}
\author{Jokin Ezenarro\fnref{fn2}}
\ead{jokin@food.ku.dk} 
\address{University of Copenhagen}
\author{Daniel Schorn-García\fnref{fn3}}
\ead{dschorng@sun.ac.za} 
\address{Stellenbosch University}
\author{Johan A. Westerhuis\fnref{fn4}}
\ead{j.a.westerhuis@uva.nl} 
\address{University of Amsterdam }

\fntext[fn1]{Centro de Investigación en Tecnologías de la Información y las Comunicaciones, University of Granada, C/ Periodista Daniel Saucedo Aranda s/n 18071, Granada, Spain.}
\fntext[fn1b]{Datharsis, Granada, Spain.}
\fntext[fn2]{Department of Food Science, Faculty of Science, University of Copenhagen, Rolighedsvej 26, Frederiksberg C 1958, Denmark.}
\fntext[fn3]{South African Grape and Wine Research Institute (SAGWRI), Department of Viticulture and Oenology, Stellenbosch University, 7600, South Africa.}
\fntext[fn4]{Biosystems Data Analysis, Swammerdam Institute for Life Sciences, University of Amsterdam. Amsterdam, Netherlands} 
\cortext[cor1]{Corresponding author} 

\date{}

\begin{abstract}
ANOVA Simultaneous Component Analysis (ASCA) is the current state-of-the-art chemometric tool for analyzing and interpreting high-dimensional experimental data from a Design of Experiment (DoE). Being a multivariate extension of the ANOVA, ASCA makes a perfect tandem with DoE. This tutorial review recommends best practices for using ASCA, building upon the long-established combination of ANOVA and DoE theory developed over the last century. These recommendations are grounded in a comprehensive literature review and illustrated through a guiding example.
\end{abstract} 
\begin{keyword}
 ANOVA Simultaneous Component Analysis \sep Permutation Testing  \sep Design of Experiments \sep high-dimensional data{\sep {Mixed Linear Models}}{\sep {Longitudinal analysis}}
\end{keyword}
\end{frontmatter}   

\section{Introduction}  \label{Introduction}

Experimental science has experienced a paradigm shift towards collecting massive volumes of data to develop and test sophisticated scientific hypotheses~\cite{sanchez2018integrated}. Often breakthroughs in science are enabled through the development of new scientific equipment~\cite{venter2001sequence,stuart2019integrative,camacho2025single}. Yet, advanced analytical instruments pose new challenges for data  analysis, including  multidimensional and multimodal data organization, {varying} signal-to-noise ratio along the measurement range, and complex reproducibility phenomena such as batch effects. Coping with these challenges requires a careful study design.

The theory behind statistical Design of Experiments (DoE) coupled with the ANalysis Of VAriance (ANOVA) of the experimental results dates back to the pioneering work of Ronald A. Fisher at the beginning of the XX century~\cite{fisher1992arrangement,fisher1936design}, and was further developed by renowned statisticians like George Box~\cite{box1978statistics} and Genichi Taguchi~\cite{taguchi1978off}, among many others. As of today, there is a well-established and extensive theory on the topic, ranging from simple (yet very powerful) concepts to more complex statistical inferential approaches~\cite{montgomery2020design}. A cornerstone of this methodology is that choices made on the experimental design may have far-reaching consequences on the statistical power, that is, the ability to find statistically significant patterns during the analysis. This visionary idea of coupling design and analysis has become the foundation for decades of clinical research~\cite{brown1980crossover}, ecology~\cite{scheiner2001design} and the -ometrics sciences, including chemometrics~\cite{esbensen2002multivariate,deming1993experimental}. When dealing with data coming from modern analytical equipment, a proper DoE becomes even more important to carry out successful studies.

ANOVA Simultaneous Component Analysis (ASCA) \cite{smilde_anova-simultaneous_2005} has recently become the state-of-the-art chemometric tool for the analysis and interpretation of multivariate and multi-factorial experimental data coming from a DoE. The term multivariate is used here to designate experiments where several responses (often in the hundreds, thousands or more) are measured in each experimental unit. Good examples are clinical, agri-food and ecological studies based on high-throughput analytical instruments, like in the areas of genomics, transcriptomics, proteomics or metabolomics. ASCA is also broadly applicable to many high-dimensional analytical data, including spectroscopy \cite{Rust2021}, hyphenated separations \cite{DAlessandro2021}, hyperspectral and chemical imaging \cite{Ezenarro2025HSI}, sensor arrays, and electronic nose or tongue systems \cite{Strani2022}. The term multi-factorial designates the situation when more than one factor is considered in the experiment. In the context of ANOVA, multi-factorial is often also termed multiway, which highlights an interesting connection between some forms of experimental designs and multiway data~\cite{smilde2005multi}. {Although ASCA was originally developed for data arising from statistical experimental design, it has only recently begun to be applied to observational data \citep{villarargaiz2022heat, camacho2024netmob, thiel2025development}. }In summary, ASCA is {es}specially powerful to unravel the complex connection between several inputs (factors and their interactions) and several outputs (the experimental responses or end-points) in the face of limited samples.

Being a multivariate extension of the ANOVA, ASCA makes a perfect tandem with DoE. 
When coupling DoE and ASCA, the experimenter can make decisions before the study is conducted, which will be instrumental to maximize the chances to find interpretable results at the end of the pipeline. A solid understanding of the interplay of the whole experimental pipeline, from design and experimentation to analysis and interpretation, is paramount to make the most of modern experimental instrumentation.

 There have been previous reviews on ASCA~\cite{bertinetto_anova_2020,smilde2025analysis}. In comparison, this tutorial review makes special emphasis in the combination of DoE and ASCA, including instrumental concepts related to statistical power, inference and data interpretation. The tutorial is designed so that practitioners with diverse technical background can make the most of it, by placing more theoretical derivations in the appendices. The rest of the paper is organized as follows: Most common decisions made during the study design are discussed in Section \ref{DoE}. Section \ref{Factorization} discusses univariate ANOVA, cornerstone of the ASCA approach. Section \ref{Pipeline} introduces the ASCA pipeline. A brief discussion on related models and alternative software tools is performed in Section \ref{Related}. Finally, Section \ref{Conclusions} concludes the work. Main sections are also accompanied with a leading example on response to treatment in breast cancer.    

 \begin{tcolorbox}[breakable,title={BOX 0: The Choice of Naming Convention},colframe=green!50!black]
 {Technical naming is a contentious issue within both ANOVA and ASCA literature, which creates some confusion. While we aim for maximum clarity, it is perhaps impossible to adopt a naming convention that satisfies every school of thought. As with other theoretical considerations in this paper, we follow the convention established by Montgomery~\cite{montgomery2020design}, who uses 'ANOVA' as an overarching framework for variance factorization and hypothesis testing in multi-factor/multi-level designs.
In this context, the term ANOVA is used broadly to encompass:
\begin{itemize}
    \item {Linear} models with fixed and/or random effects;
    \item Alternative parameter estimation approaches via averaging, least squares, or maximum likelihood;
    \item Different approaches to inference, from parametric to nonparametric solutions.
\end{itemize}
This is the most straightforward approach, while not universally accepted. ASCA would also benefit of a similar naming strategy, thereby reinforcing the sense of a unified methodology derived from its parent framework, ANOVA. }
 
 \end{tcolorbox}

\section{The DoE pipeline}  \label{DoE}

The experimental design follows a number of steps~\cite{montgomery2020design} with relevant choices regarding the study details. If the adequate choices are made, the statistical power of the analysis will be optimal. The statistical power is a central concept in this tutorial: it refers to the probability that the analysis will detect a true effect in the responses. Refer to Box 1 for a number of useful definitions for the rest of the manuscript.

\begin{tcolorbox}[breakable,title={BOX 1: Definitions},colframe=green!50!black]
\textbf{Response (or Response Variable)}: The measured outcome of the experimental study.
\\ \textbf{Experiment (or Experimental Run)}: A single measurement of the response(s). 
\\ \textbf{Factor}: A variable that is manipulated at (controllable factor) or takes (uncontrollable factor) different levels through the study that may cause a change (an effect) in the response(s).
\\ \textbf{Interaction}: The additional synergistic or antagonistic effect of the combination of several factors in the response(s).
\\ \textbf{Term}: A component of the model, either an individual factor or interaction, used to describe the relationship between factors and response(s).
\\ \textbf{Level}: Each of the different values of a factor considered in the study.
\\ \textbf{Cell}: Group of experiments made for a combination of levels of all factors.
\\ \textbf{Treatment}: Experimental condition or intervention applied through the experimentation. This often refers to the combination of levels of the design factors.
\\ \textbf{Experimental Unit}: The smallest unit to which a treatment is applied independently.
\\ \textbf{Repeated Measures}: Multiple experiments for the same experimental unit under different conditions or over time.
\\ \textbf{Replicates}: Independent experimental units assigned to the same treatment.
\\ \textbf{Pseudoreplicates}: Replicates that are not truly independent but rather belong to the same experimental unit.
\\ \textbf{Fixed Factor}: A factor whose specific levels are the focus of the study, the conclusions apply only to those and are not expected to generalize to other possible levels.
\\ \textbf{Random Factor}: A factor whose levels are randomly sampled from a larger population. The goal is to generalize the conclusions to the entire population of levels.
\\ \textbf{Crossed Relationship}: Factors are crossed when every level of one factor occurs in combination with every level of the other factor.
\\ \textbf{Nested Relationship}: Factors are nested (or hierarchical) when the levels of one factor are unique to, and do not repeat across, the levels of another factor. 
\\ \textbf{Order of a Term}: The order of a term represents its position in a hierarchical model. Main factors have order 1. The order of nested factors is the order of the factor to which they are nested plus 1. The order of the interactions is the sum of the orders of the factors considered in it.
\\ \textbf{Randomization}: Randomly assigning levels to the experimental units to break any systematic association between (often unknown) confounding variables and the levels. 
\\ \textbf{Blocking}: Organizing the experimental units into homogeneous groups with respect to some external factor that may influence the outcome. 
\\ \textbf{Null hypothesis}: The default statement that assumes no effect of {a} model term {on} the response(s).
\\ \textbf{Alternative hypothesis}: The position that rejects the null hypothesis.
\\ \textbf{Type I Error (TIE)}: The error of rejecting the null hypothesis when it is actually true. It is also known as a "false positive". The TIE may also refer to the probability or rate of false positives.
\\ \textbf{Type II Error (TIIE)}: The error of failing to reject the null hypothesis when it is actually false. It is also known as a "false negative". The TIIE may also refer to the probability or rate of false negatives.
\\ \textbf{Significance Level}: The probability threshold used to determine whether a statistical result is significant. It is the maximum acceptable TIE rate.
\\ \textbf{Statistical Power {($1-$TIIE)}}: The probability of correctly rejecting the null hypothesis when it is false. It is the ability of a test to detect an effect of a given size if it truly exists\footnote{{Please note that while the terms TIE and $1-$TIIE are often denoted as $\alpha$ and $\beta$, we avoid this notation in this paper to prevent confusion with the ANOVA parameters.}}. 
\end{tcolorbox}

Choices in the experimental design obviously require a strong expertise on the research area of the study, but also a deep understanding on data science. This understanding is necessary, firstly, due to the inherent complexities in making the most of the data coming from modern analytical instruments (e.g., unreliable instruments, batch/technical effects, etc.). Secondly, 
statistical theory shows that some DoE decisions will maximize statistical power in specific ANOVA analyses: well-known examples are that repeated measures designs are  optimal for modeling within-subject variance, and that factorial experiments make the most efficient use of the data and are optimal to infer interactions~\cite{montgomery2020design}. Finally, more subtle relationships between DoE choices and statistical power for some given data characteristics (like data imbalance, missing data patters, or complex distributions) can be inferred with computational means~\cite{camacho_population_2024}. Thus, it is highly recommended that the experimental design is performed by multidisciplinary teams that include the relevant expertises. As Fisher claimed in the first Indian Statistical Congress in 1938 \textit{"To consult the statistician after an experiment is finished is often merely to ask him to conduct a Post mortem examination. He can perhaps say what the experiment died of."}~\cite{fisher1938}. This also reflects the fact that scientific research is ultimately based on data, and understanding how data tends to behave is paramount to make the right decisions \textit{a priori}. 

A list of steps and corresponding choices in the experimental design are depicted in Figure \ref{fig:DoEPipeline} and described below:

\begin{figure}
    \centering
    \includegraphics[width=1\linewidth]{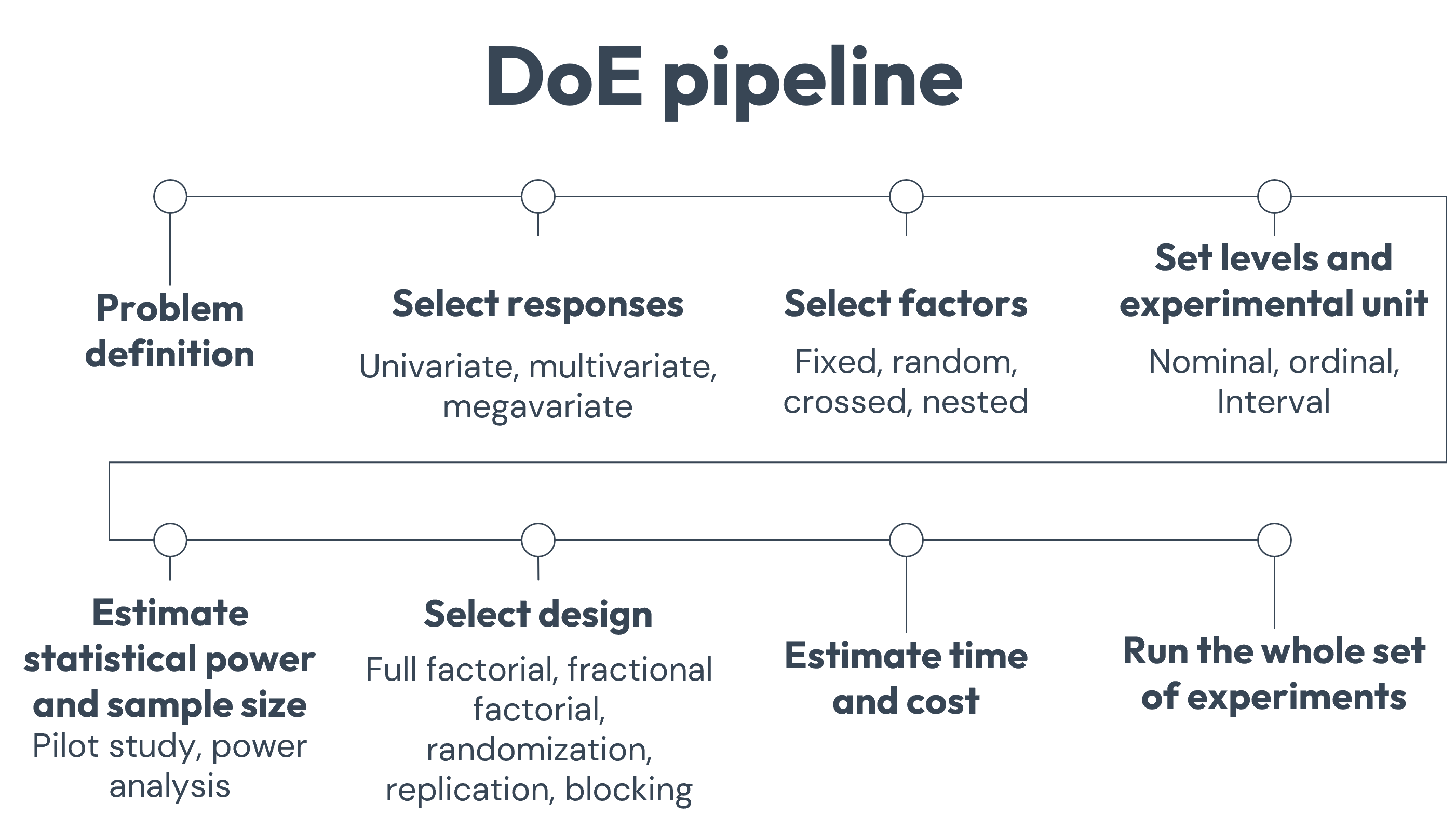}
    \caption{Illustration of the DoE pipeline: steps to follow prior to the study.}
    \label{fig:DoEPipeline}
\end{figure}

\begin{itemize}

\item The first step is to \textbf{provide an accurate definition of the aim of the study}. One example could be to test the hypothesis: {i}n patients of a given condition, their metabolite levels contain information to predict whether a given treatment is going to be effective. An accurate definition of the goal and the scope of the study will help clarify many of the subsequent decisions: What part of the metabolism are we considering? Where and how are we measuring it? What patients will participate in the study? What molecular variant of the disease are we focusing on? Clarifying all relevant questions will allow us to narrow down the study to a practical definition.

\item \textbf{Select the set of responses} to test the hypothesis, in the range from a single univariate response in traditional experiments, to the megavariate case in omics sciences. Following our previous example, metabolomics assays of liquid biopsies may be a potential candidate for the study. Clinical experts and analytical chemists may identify the most suitable analytical technology. It should be noted that while multivariate responses are more complex and require more sophisticated data analysis tools~\cite{anderson_permanova_2013}, there is a potential advantage in statistical power associated to the consideration of multivariate responses~\cite{camacho_variable-selection_nodate}. Furthermore, considering multivariate responses is more consistent to our understanding of biochemical processes, where several compounds need to be jointly present to trigger them.

\item \textbf{Select the set of factors} under which the experiment is performed, including their nature (fixed or random), and their relationships (crossed or nested), see BOX 1 for definitions.

\item \textbf{Select the levels and the experimental unit}. For a qualitative factor, levels do not have an ordered relationship, and the factor will be treated as purely categorical {(nominal). Quantitative factors can be directly treated as {continuous variables} or} we can select a set of reasonable levels, and the factor may be treated as nominal or {continuous} in the analysis, with different interpretation. A quantitative factor may also be broken down in intervals, like transforming age into 'children', 'adult' and 'elderly', but it should be properly justified. Proper selection of levels and boundaries is crucial for accurate partitioning and valid interpretation. Levels are also connected to the definition of the experimental unit, which is crucial for accurate inference.

\item \textbf{Estimate statistical power and sample size}. If possible, run some experiments to understand time constraints and practicalities of the experiments. This may not only provide important information on what to expect in the complete study, but also allow us to detect potential pitfalls and get useful information on expected ranges and correlations of the responses, number of missing values, distributions, etc. Such a small \textbf{pilot study} should at least give information on reproducibility, suitable levels of the factors and effect size, to be able to provide a better power calculation for the full study. A power curve is a standard tool to estimate the minimum sampling size that provides a reasonable statistical power for the study. 
In standard univariate ANOVA, or when relatively few responses are considered, it is possible to use analytical methods to derive power curves based on typical assumptions (such as normality)~\cite{hickey2018statistical}. However, in the context of more than just a few responses, or to assess the violation of any possible statistical assumptions, numerical methods \cite{anderson2003permutation} are a viable alternative. Simulated power curves~\cite{camacho_population_2024} allow us to consider inherent complexities in the data, while assessing not only the influence of the sample size, but also of the relative number of levels of all factors, on the statistical power. The information gained when running some initial experiments in the previous step can be leveraged into the simulated power curve to derive more accurate estimates o{f} the statistical power.

\item \textbf{Select the structure of the design} (e.g., full factorial, fractional, other) and practical implementation (randomization, blocking and replication). 
Randomization and blocking improve the statistical power in front of uncontrolled and controlled confounding variability, respectively. Replication is key for estimating the inherent variability of the system under study, and accurately identifying statistically significant patterns. 

\item \textbf{Estimate the time and cost of the complete study}. If several alternatives for the study are considered, pick the one that maximizes statistical power for the hypothesis of interests while minimizing time and/or cost.

\item \textbf{Carry out the study}.

\end{itemize}

In Box 2 we introduce the guiding example of this tutorial, where the previous steps of DoE are discussed.

\begin{figure}
    \centering
    \includegraphics[width=0.75\linewidth]{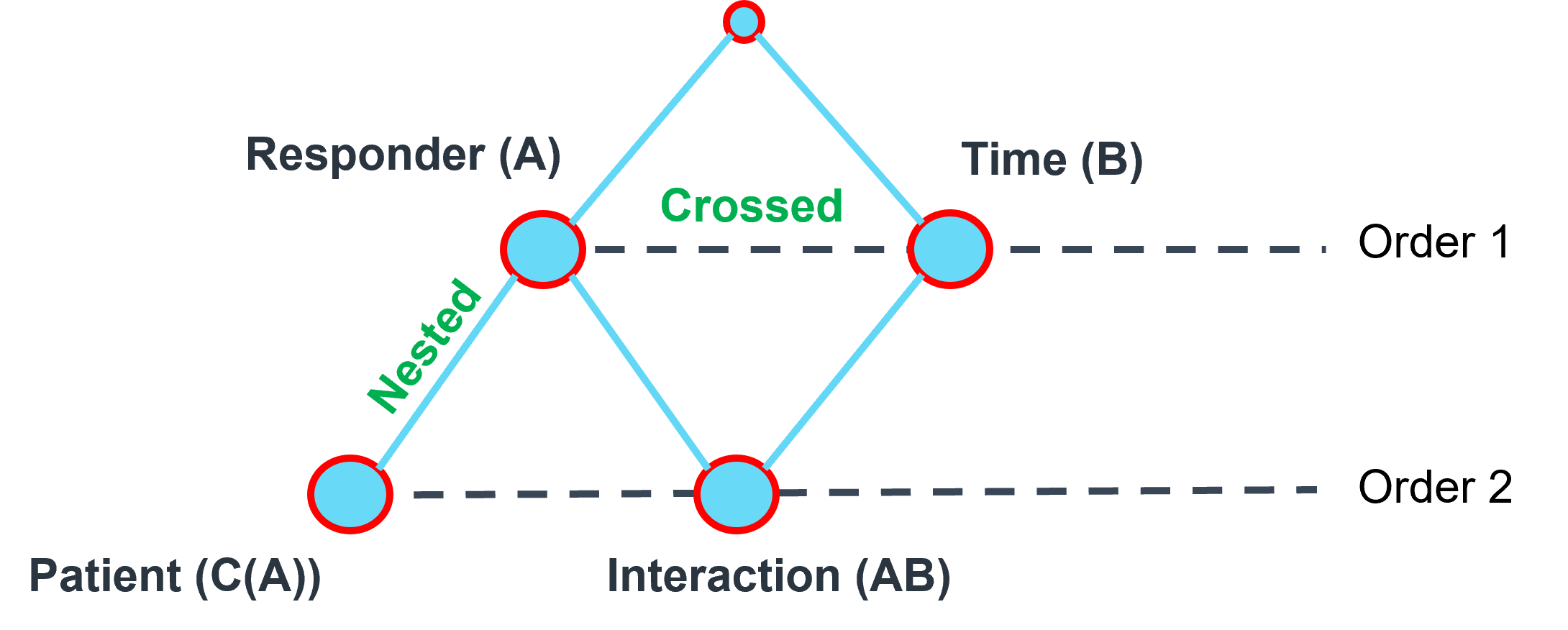}
    \caption{Hasse diagram of the example in BOX 2.}
    \label{fig:Hasse}
\end{figure}

\begin{tcolorbox}[breakable,title={BOX 2: Guiding example: Response to treatment in breast cancer},colframe=green!50!black]
The guiding example of the paper corresponds to a prospective study on treatment response in breast cancer with untargeted metabolomics~\cite{diaz2022predicting}. The treatment considered is the neoadjuvant chemotherapy (NACT): a course of chemotherapy administered before surgery. NACT is a popular treatment for breast cancer, but unfortunately not all patients benefit from it, and therefore it is critical to differentiate between the subjects that will respond positively and those who will not, in order to choose alternative and more effective therapies. Although this differentiation can only be done after the study,  the metabolites linked to this Responder factor are expected to have a large contribution in its effect matrix. In particular, we focus on the triple negative (TN) molecular variant of breast cancer: patients who neither express hormone receptors of estrogen (ER) and progesterone (PR) nor human epidermal growth factor 2 (HER2). Venous blood samples were collected under fasting conditions at three different time points: basal, presurgery and postsurgery. Pathological response was assessed by histopathological analysis of the samples obtained during surgery, using the five-step scale Miller and Payne (MP) grade based on reduction in malignant cellularity after the complete treatment, from no reduction (MP1) to no malignant cells remaining (MP5). Those five levels were transformed into a nonresponder {(NR)} group (for MP grades 1-3) and a responder {(R)} group (for MP grades 4-5), a division meaningfully connected to the prognostic potential. The study considers a total of 21 patients (with 63 blood samples), with 13 in the group of responders and 8 nonresponders. A non-target metabolomics analysis with Liquid Chromatography coupled with Mass Spectrometry (LCMS) was performed on the data along with quality controls. After peak detection, alignment, and data filtering a data matrix with 63 rows and 112 columns (corresponding to metabolites) was obtained.

Having the previous example in mind, let us go through all the steps in the design of the study:

\begin{itemize}

\item The aim of the study is to test the hypothesis: "In TN breast cancer patients, the metabolome of blood samples provides information to predict if NACT is going to be effective."  

\item \textbf{Select the set of responses}: We use LCMS as the analytical technology and a total of 112 responses (metabolites) are considered.   

\item \textbf{Select the set of factors}: Our main Factor is `Responder'. We also have a Factor `Time', with the repeated measures for each patient, and a Factor `Patient'. Factors 'Responder' and 'Time' are fixed (statistical inference is limited to {the} levels {that} are considered) and crossed (all combinations of 'Responder' and 'Time' are included in the design). Factor 'Patient' is random, because its inference is meant to generalize to the population of patients, and not just to the ones in the study. Besides, it is nested in Factor 'Responder', because each patient can only belong to a single level of the Responder factor. A Hasse diagram \cite{marini2015analysis} with the order and relationships of the factors is presented in Figure \ref{fig:Hasse}.

\item \textbf{Select the levels and experimental unit}: The levels of Factor `Responder' are two: responders and non-responders. The levels of Factor `Time' correspond to the times where repeated samples are taken for each patient: basal, presurgery and postsurgery. In total, we have 21 patients (levels in Factor `Patient'). Each patient is an experimental unit.

\item No previous experiments were run for this case. No power analysis was conducted prior to the study. The sample size was limited by the rate of TN patients treated in the hospital where the study was conducted. 

\item \textbf{Select the structure of the design}: the design is a repeated measures study, with a full factorial structure between the patients and the time measurements which blocks the measurements in time. The fact that we have repeated measures is instrumental to model the individual `Patient' and improve the statistical power of the study. Given the prospective nature of the study, randomization of the assignment of patients to the group of responders was not possible. The replication is carried out with the sample size, 21 patients. 

\end{itemize}

\end{tcolorbox}

\section{Analysis of Variance}  
\label{Factorization}

The theory of ANOVA applied over a single response variable has been developed throughout the last century, and as such it should be the preferred reference for devising good practices in the younger ASCA. For this reason, we devote this section to ANOVA concepts that are principal to ASCA.

Figure \ref{fig:DiagPipeline} shows the pipeline of steps being used for ANOVA and ASCA, which starts after completion of the planned experiments. Before the modeling takes place, it is common to explore the data to search for anomalies, missing data, etc. After removal of these problematic issues, the modeling process starts. The experimental setup used mixes the effects of the different factors, and the role of ANOVA is to unmix these effects and estimate the proper effect sizes \cite{fisher1925}. The distributional assumptions of ANOVA with respect to normality and equal variance apply to the residuals of the model, and the measured data is preprocessed to attain these assumptions. Thus the initial steps up to factorization are {adapted and iterated} until the assumptions o{n} the residuals are fulfilled. These steps differ between ANOVA and ASCA due to the multivariate character of the data in the latter method. Finally, the statistical inference of the model parameters can be assessed and their conclusions can be complemented with a posthoc test.

In this ANOVA section we focus mainly on the part of the pipeline from the factorization onwards. The data curation and pretreatment steps will be discussed in the ASCA section.

\subsection{Factorization}

\subsubsection{Means model}
The {simplest} ANOVA model is for the balanced situation where for each of the $G$ levels for a single factor, $N$ subjects have been measured. In total there are $I = G*N$ measurements $y_{gn}$ for a single response variable. The Means model for this situation is
\begin{equation}
y_{gn} = \mu_g + \epsilon_{gn}
\label{eq:Anova_means1}
\end{equation}
The level effect for $g$, $\mu_g$, can be simply obtained by calculating the mean for each level $g$. It is however not possible to use this approach for more complex ANOVA models and thus usually a more general model is used to describe the variation in the data.
\begin{equation}
y_{gn} = \mu + \alpha_g + \epsilon_{gn}
\label{eq:Anova_means2}
\end{equation}
Now $\mu$ is the overall mean and $\alpha_g$ is the additional effect for level $g$. This is often called the effects model~\cite{montgomery2020design}. This model has $G + 1$ parameters to estimate for only $G$ levels in the factor; thus, it is over-parameterized. This would lead to problems in estimating the parameters as there are infinite solutions to the problem. The most common approach to deal with this is to calculate the level effects $\alpha_g$ in such a way that $\sum_{g=1}^{G} \alpha_g = 0$. The model parameters $\mu$ and $\alpha_g$ can now be estimated by decomposing all sources of variation for each measurement.
\begin{equation}
y_{gn} = \bar{y} + (\bar{y}_g - \bar{y}) + (\bar{y}_{gn} - \bar{y}_g)
\label{eq:Anova_means3}
\end{equation}
Where $\bar{y}$ is the estimate for $\mu$ and calculated as the overall mean, and $\bar{y}_g$  is the mean of all subjects in group $g$ and thus $(\bar{y}_g - \bar{y})$  is the additional effect of level $g$.  $(\bar{y}_{gn} - \bar{y}_g)$ is the estimate of the residual.

In an ANOVA model the total variation in the measurement $y$ over all levels and all subjects is thus decomposed in a part due to the groups and a part within the groups. The variation is calculated as the sum of squares of each of these sources.

\begin{equation}
\sum_{g=1} ^G \sum_{n=1} ^N (y_{gn} - \bar{y})^2 = \sum_{g=1} ^G \sum_{n=1} ^N (\bar{y}_g - \bar{y})^2 + \sum_{g=1} ^G \sum_{n=1} ^N (y_{gn} - \bar{y}_g)^2
\label{eq:Anova_means4}
\end{equation}
The Total sum of squares $(SS_T)$ is decomposed into the sum of squares due to the $G$ levels $(SS_G)$ and the residual sum of squares $(SS_E)$.

In the statistical test to see whether the variation between the groups is large compared to the variation within a group, the sum of squares have to be corrected for the number of independent contributions to their sum of squares, which are called the degrees of freedom (DoFs). The variation due to $G$ groups has $G-1$ independent parts due to the sum-to-zero constraint of the group levels. The total sum of squares has $I-1$ independent parts as the mean was already estimated and used to calculate the $SS_T$. For the error sum of squares, the $G$ group means were used and thus $I - G$ independent contributions are left. The Mean Squares (MS) are obtained by dividing the sum of squares by the corresponding DoFs. Finally, the test for significant group differences is performed by comparing the Groups Mean Square with the Residual mean Square. This test is called the F-test, as its result follows an F-distribution if there are no group differences. If the calculated F-value is larger than what can {be} expected by chance (e.g. 95$\%$ confidence) the result is called statistically significant.

\subsubsection{Regression model}
Another approach to {estimate the parameters of} the ANOVA model is by using a regression model.
{
\begin{equation}
\mathbf{y}_g = \mathbf{1}_{n_g} \mu + \mathbf{1}_{n_g} \alpha_g + \boldsymbol \epsilon_g
\label{eq:Anova_regression1}
\end{equation}}
In this model {$\mathbf{y}_g$} is a vector with all $y_{gn}$ values, {$\mathbf{1}_{n_g}$ represents a column vector of ones of the size of $\mathbf{y}_g$} and $\alpha_g$ represents again the additional effect of group $g$ on top of the $\mu$ value. {Finally, $\boldsymbol{\epsilon}_g$ is the vector with residuals for the samples in group $g$.} This is a Linear Model and it assumes that the residuals are independent and come from the same distribution. In most cases a normal (also called Gaussian) distribution is assumed for the residuals, and then the model is often referred to as General Linear Model {and estimated through Ordinary Least Squares (OLS)}. If the residuals are described by another predefined distribution, then such a model is often called a Generalized Linear Model. Unfortunately, the acronym of both terms is GLM, which creates some confusion in the literature.

As explained above, the model is over-parameterized, and one solution is to constrain the estimation of the $\alpha_g$ values such that they sum to 0. For a 3 level example this could be written as $\alpha_1 = {0} -\alpha_2 -\alpha_3$. Then, when we write the equation { for each group as follows:}
{
\begin{gather}
\mathbf{y}_1 = \mathbf{1}_{n_1} \mu - \mathbf{1}_{n_1} \mathbf{\alpha}_2 - \mathbf{1}_{1}\mathbf{\alpha}_3 + \boldsymbol \epsilon_1 \\ \nonumber
\mathbf{y}_2 = \mathbf{1}_{n_2}\mathbf{\mu} + \mathbf{1}_{n_2}\mathbf{\alpha}_2  + \boldsymbol{\epsilon}_2 \\ \nonumber
\mathbf{y}_3 = \mathbf{1}_{n_3}\mathbf{\mu} + \mathbf{1}_{n_3}\mathbf{\alpha}_3 + \boldsymbol{\epsilon}_3 \\ \nonumber
\label{eq:Anova_regression2}
\end{gather}}
where $\mathbf{y}_1, \mathbf{x}_{\mu_1}, \mathbf{x}_{2_1}$and $\boldsymbol{\epsilon}_1$ only refer to the replicates of group $1$, etc. 

%

 To use the matrix form, it becomes clear that the model matrix $\mathbf{X}$ should have 3 columns, one for indicating $\mathbf{x}_\mu$, $\mathbf{x}_2$ and $\mathbf{x}_3$. A column for $\mathbf{x}_1$ is not needed as $\alpha_1$ does not appear in any of the equations. The $\alpha_1$ value,  however, can always be calculated as $\alpha_1 = 0 - \alpha_2 - \alpha_3$. The { model matrix $\mathbf{X}$, combining the columns for $\mathbf{x}_\mu$, $\mathbf{x}_2$ and $\mathbf{x}_3$ is presented in Table \ref{Sum Coding matrix}}. 

\begin{table}[h]
\caption{{Sum-coded model matrix} $\mathbf{X}$}
    \label{Sum Coding matrix}
    \centering 
    \begin{tabular}{llll}
        \toprule
        \textbf{Group} & $\mathbf{x}_\mu$ & $\mathbf{x}_2$ & $\mathbf{x}_3$ \\
        \midrule
        Group 1 samples & 1 & -1 & -1 \\
        Group 2 samples & 1 & 1 & 0 \\
        Group 3 samples & 1 & 0 & 1 \\
        \bottomrule
    \end{tabular}
 \end{table}
This way of coding {the model matrix} $\mathbf{X}$ is called Sum-coding or Deviation-coding and was used by Thiel \textit{et al}. \cite{thiel_asca_2017} in their ASCA+ paper. There is no column for $\mathbf{x}_1$ to estimate $\alpha_1$. The other parameters, {$\boldsymbol \theta = [\mu,\alpha_2, \alpha_3]^T$, can be estimated as follows:
\begin{equation}
\hat{\boldsymbol  \theta} = (\mathbf{X}^T\mathbf{X})^{-1}\mathbf{X}^T \mathbf{y}
\label{eq:Anova_regression3}
\end{equation}}
The advantage of the regression model estimation of {$\boldsymbol \theta$} becomes clear in unbalanced situations where the number of subject{s} per group is not the same. Thiel \textit{et al}. \cite{thiel_asca_2017} show a clear example of this situation. There are also other approaches to solve the over-parametrization of the model, like reference and weighted coding (see Appendix A).

\subsubsection{Two-way ANOVA}
For designs with two or more crossed factors, the situation becomes more complicated as there could be an interaction between the two factors (e.g., $A$ and $B$) each with different number of levels (e.g., $A$ with 3 levels and $B$ with 2 levels). The interaction effect (usually indicated as $AB$) is a correction for the synergy or antagony if the effect of the combination of the two factors is different from the sum of their main effects. {The two factor model with interaction can be written as follows:
\begin{equation}
\mathbf{y} = \mathbf{x}_\mu \mu + \mathbf{X}_{A} \boldsymbol \alpha + \mathbf{x}_{B} \beta + \mathbf{X}_{AB} (\boldsymbol  \alpha \boldsymbol \beta) + {\boldsymbol \epsilon}
\label{eq:Anova_twoway}
\end{equation}
In this equation $\mathbf{x}_{B} \beta$ represents the additional effect of Factor $B$, and $\mathbf{X}_{AB} (\boldsymbol{\alpha\beta})$ represents the additional effect of the interaction between Factors $A$ and $B$. As Factor $B$ only has two levels, a single column in the model matrix $\mathbf{X}$ is sufficient.} In this example we use the Sum coding again with $\alpha_1 = 0 -\alpha_2 - \alpha_3$ and $\beta_1 = 0 - \beta_2$.



The combined model matrix for the two factor model is in Table \ref{Two_factor_Sum_Coding_matrix}. The coding columns for the interactions {in the model matrix $\mathbf{X}$} are formed by multiplying each column for Factor $A$ with each column for Factor $B$ in an element wise manner. The same approach can be used for Reference coding.

\begin{table}[h]
\caption{Two factor Sum Coded model matrix $\mathbf{X}$}
    \label{Two_factor_Sum_Coding_matrix}
    \centering 
    \begin{tabular}{lllllll}
        \toprule
        \textbf{AB Group} & $\mu$ & $\mathbf{x}_{A2}$ & $\mathbf{x}_{A3}$ & $\mathbf{x}_{B2}$ & $\mathbf{x}_{A2B2}$ & $\mathbf{x}_{A3B2}$ \\
        \midrule
        $A_1B_1$ samples & 1 & -1 & -1 & -1 &  1 &  1 \\
        $A_2B_1$ samples & 1 &  1 &  0 & -1 & -1 &  0 \\
        $A_3B_1$ samples & 1 &  0 &  1 & -1 &  0 & -1 \\
        $A_1B_2$ samples & 1 & -1 & -1 &  1 & -1 & -1 \\
        $A_2B_2$ samples & 1 &  1 &  0 &  1 &  1 &  0 \\
        $A_3B_2$ samples & 1 &  0 &  1 &  1 &  0 &  1 \\
        \bottomrule
    \end{tabular}
 \end{table}

\begin{figure}
    \centering
    \includegraphics[width=1\linewidth]{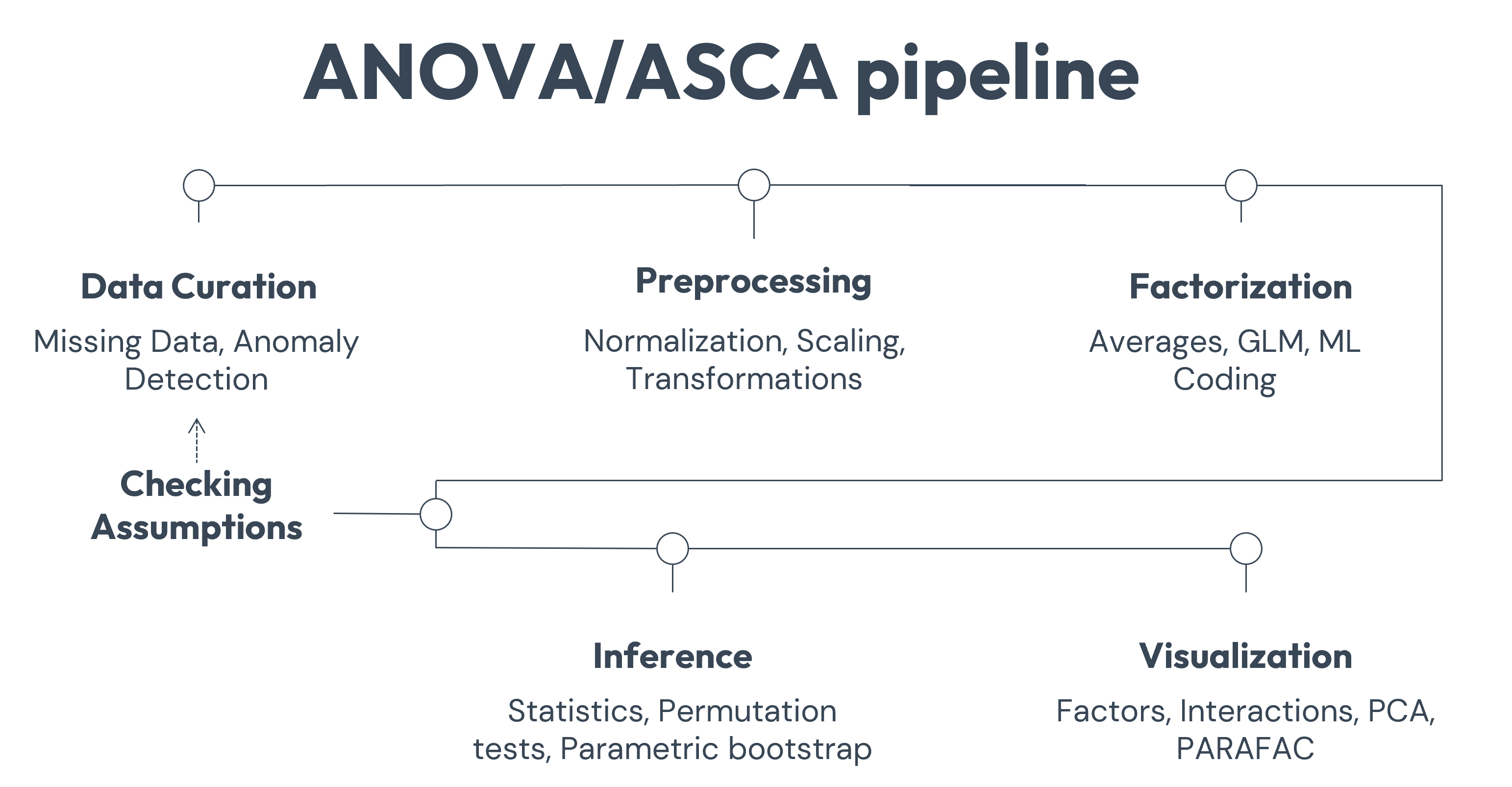}
    \caption{Illustration of the analysis (ANOVA/ASCA) pipeline.}
    \label{fig:DiagPipeline}
\end{figure}

Up till now we have assumed that the number of replicates in each group is equal. In that case any column of Factor $A$ is orthogonal to any column of Factor $B$ and also to any column of the interactions. This means that the effect of Factor $A$ can  be estimated independently of Factor $B$ and also of the interaction between $A$ and $B$. In this case their effects are independent and their estimation is not affected by the other factors.

In the unbalanced situation where the number of replicates is not equal for all groups the columns of the coding matrix lose their orthogonality. The estimated effects of all factors and their interactions become dependent on each other{. This effect is often negligible when unbalance is low. When not, a} common approach to make the data balanced again is by removing samples or by {imputing} samples, both {of} which are not optimal. A recently introduced weighted coding \cite{Ali2020WEASCA} can improve the situation, see Appendix A.

A direct consequence of an unbalanced design and the non-orthogonality of the factors is that the sum of the SS of the model terms and residuals do not necessarily add up to the total. This reflects double-counting of variance in the factorization and in the effect size estimates. The more double-counting, the less reliable these estimates and associated statistics are. This can be mitigated in different ways{, and the most extended approach is to resort to Type I, Type II and Type III SSs\footnote{Do not confuse Type I and II errors, as defined in Box 1, with Type I, II and III SSs, which are totally different concepts.}. Unfortunately, there is a significant disagreement regarding which approach is better and how it should even be computed, {as discussed in Appendix B.}

\subsubsection{Nested factors}
In some designs there is a hierarchy in the factors where some levels of the nested factor are only appearing in one level of the higher factor. For instance, in the example of BOX 2, Factor $C(A)$ 'Patient' is nested {in} Factor $A$ 'Responder', because each patient can be a responder or non-responder, but not both. Because not all levels of the nested factor are present for all levels of the higher factor, {the interaction between Factor $A$ and Factor $C(A)$ does not exist, leading to the following model, where $\gamma_{c(a)}$ represents the effect of patient {in} $c$}. 

\begin{equation}
y_{abc} = \mu + \alpha_a + {\beta_b + \gamma_{c(a)} + (\alpha \beta)_{ab}} + \epsilon_{abc}
\label{eq:nested_model1}
\end{equation}

\subsubsection{Linear Mixed Models}
\label{sub:LinearMixedModels}

Models that combine random and fixed factors are referred to as linear mixed models (LMMs). Most often, fixed factors have levels of interest (we really want to know what is the difference between the levels of a fixed factor, that is, the fixed effects) while we are not interested in the actual levels of random factors but rather {in} their variance components. These components mainly serve to deal with the grouping of the samples taking into account their variance and covariance structure. 

It is a common misunderstanding that linear mixed models cannot be modeled with the ANOVA {estimation procedures} described before, including {averaging in the} effects model and {OLS} {in the regression model. In fact, for balanced data, the variance estimates for random effects and residuals by these models equal} those obtained with more developed approaches, like those based on the method of maximum likelihood on the normal distribution, e.g., Restricted Maximum Likelihood (REML)~\cite{montgomery2020design}. {REML represents the state of the art for the estimation of mixed models in univariate ANOVA, at the expense of computational complications, given its superior ability to handle unbalanced data~\cite{martin_limmpca_2020} and to derive confidence intervals for variance estimates and covariances between random terms~\cite{montgomery2020design}. However, its extension to multivariate data is complex and it is not clear whether multivariate maximum likelihood approximations~\cite{Madssen2021_RMASCA,martin_limmpca_2020} are indeed superior to their OLS counterparts.}
A discussion on maximum likelihood estimates for LMMs is provided in Appendix {C}.

\subsubsection{Repeated Measures}

A special case of a mixed model is a repeated measures model~\cite{Madssen2021_RMASCA}, which assumes multiple measurements for the same subject under different conditions (quite often over time). In many experimental designs the effect of different treatments are assessed at multiple time points. The measurements for a single subject are expected to have less variation than measurements of different subjects. For instance, the example of BOX 2 is a Repeated Measures design, with three repeated measures for each patient. As discussed in the previous section, repeated measures models can be estimated with any of the ANOVA estimation procedures, the effects model, {OLS} or maximum likelihood. A discussion on maximum likelihood estimates for repeated measures is provided in Appendix {C}.

\subsection{Checking Assumptions}

The validity of the ANOVA factorization and inference relies on a number of assumptions. Therefore, the analyst should check these assumptions before interpreting the results of the ANOVA table. If assumptions are not met, the ANOVA should be run again by applying some corrections in the pipeline.  

\begin{itemize}
    \item \textbf{Distribution of residuals}. The most stringent assumptions are for parametric ANOVA, which assumes a normal distribution in the residuals. The validity of that assumption can be checked by inspecting Quantile-Quantile (Q-Q) plots where the real distribution of the residuals is compared to the percentiles of a normal distribution, or by using a suitable test for normality. If the residuals deviate from a straight line in the Q-Q plot, or if the test for normality fails, then we can either choose a model with less stringent assumptions or apply data transformations (log, root squared, Box-Cox, rank) and run ANOVA again.
    The least stringent assumptions are for non-parametric tests, which are limited in flexibility, that is, in the complexity of the experimental designs we can consider. A useful tradeoff is permutation testing~\cite{anderson2003permutation}, a distributional-free (semi-parametric{~\cite{anderson2005permutational}}) approach that is robust for a large set of distributions.  

    \item \textbf{Independence of residuals}. Independence remains a critical assumption because ANOVA relies on the experimental design to define the structure of factor effects. Residuals must be independent across experimental units; otherwise, correlations among pseudoreplicates (for example, between repeated measures or nested samples) may inflate the apparent significance of effects. Independence can only be ensured through proper design, not through post-hoc corrections. A common solution to handle pseudoreplicates is to take the average among them, so that both variance estimates and DoFs are corrected. {More developed techniques can be found~\cite{aarts2014solution}.}

    \item \textbf{Homoscedasticity (equal variance)}. ANOVA assumes that the residual variance is comparable across cells. Violations can be examined by plotting box-plots of residuals grouped by factor levels~\cite{montgomery2020design}, or by inspecting the residual variance matrix for heterogeneity. Strong differences in residual spread may indicate that some cells are inherently more variable, potentially influencing the magnitude of their modeled effects. Transformations and/or less stringent tests may also be a valid solution in this case. 
\end{itemize}

These assumptions do not invalidate ANOVA when mildly violated, but this should be considered in the interpretation of the model. Proper diagnostics and thoughtful design remain the best safeguards for robust inference.

\subsection{Inference}

Table \ref{tab:ANOVA} shows an example of ANOVA table, where several quantities for each term are provided in a row, so that they can be easily compared to the residuals and the totals in the last two rows. The design of the example is consistent with the Hasse diagram in Figure \ref{fig:Hasse}, with two crossed fixed Factors ($A$ and $B$) and a random Factor $C(A)$ nested in $A$. The second column represents the total Sum-of-Squares (SS) for each factor {, and the third column is its percentage of the total.} The fourth column represents the DoFs. The DoFs of a main factor is the number of levels minus one. The DoFs of an interaction is the product of the DoFs of its factors. The DoFs of a nested factor is computed as its number of levels minus the number of levels in the factor to which it is nested. The DoFs of the residuals is the total (number of observations minus 1 in the data) minus the DoFs of all terms in the model. The MS (ratio between the SS and the DoFs) is in the fifth column. The F-ratio for each term is the ratio between the MS of that term and a reference MS, which depends on the fixed/random nature of the terms and the relationships among terms~\cite{anderson2003permutation}. Finally, the \textit{p}-value is obtained for the F-ratio from the inference system (either parametric, distributional free or non-parametric). 

To interpret the Table, often one first identifies the statistically significant terms (those which \textit{p}-values are below the significance level, in this case only Factor $B$) and then interpret their practical importance by looking at the relative effect sizes. Different estimates of the effect sizes are possible~\cite{olejnik2003generalized}, and the $\%SS$ and the MS are useful indicators.

\begin{table}[h]
    \caption{Example of multi-factorial ANOVA table. The table includes the statistical decomposition and inference information, following a traditional format that includes: the Sum of Squares (SS) associated with each term (factors and interactions), its percentage of the total sum of squares (\%SS), the number of Degrees of Freedom (DoFs), the mean sum of squares (MS), the F-ratio (F) and the \textit{p}-value. }
     \label{tab:ANOVA}
    \centering 
    \begin{tabular}{llllllll}
 & SS & $\%SS$ & DoFs & MS & F & \textit{p}-value \\ 
 \hline 
 \hline 
    Factor $A$        & 0.76  & 1.01   & 1  & 0.76 & 0.81 & 0.37 \\ 
    Factor $B$        & 5.62  & 7.43   & 1  & 5.62 & 5.95 & 0.02 \\ 
    Factor $C(A)$     & 33.62 & 44.48  & 36 & 0.93 & 0.99 & 0.51 \\ 
    Interaction $AB$  & 1.58  & 2.09   & 1  & 1.58 & 1.67 & 0.20 \\ 
    Residuals       & 34.02 & 45.01  & 36 & 0.95 &  & \\ 
 \hline 
    Total           & 75.59 & 100.00 & 75 & 1.00 &  &  \\ 
 \hline 
 \hline 
    \end{tabular} 
\end{table}

\subsection{Post-hoc analysis}

ANOVA tests only reveal the statistical significance of a model term (either a factor or interaction), but it does not say anything about which levels or cells are different to which, and how (either in the positive or negative direction in the response(s)). Thus, once a term is found significant, additional visualizations and/or tests need to be performed to obtain more detail. We generally call these post-hoc visualizations/tests. There are several of these tests, including the Fisher's Least Significant Difference (LSD) plots and Tucker's Honestly Significant Difference (HSD) plots.

\section{The ASCA pipeline}  \label{Pipeline}
When analyzing high-dimensional (or multivariate) experimental data, similarly to univariate data, one of the first questions to address is how is the data affected by the experimental factors, this is, how much of the variation can be attributed to the factors under study. Section \ref{Factorization} showed how classical ANOVA provides a way to answer this question for a single response variable, by partitioning the variation into contributions from each factor and their interactions \cite{Hoaglin_Mosteller_Tukey_1991}. However, many datasets are multivariate by nature: spectroscopic profiles, chromatographic fingerprints, sensor arrays, or omics datasets contain hundreds or thousands of variables measured simultaneously.

For dealing with several responses in the same experiment, we can make an ANOVA test individually for each of them. The significance level in such multiple testing context is typically adjusted (corrected) in order to identify as many significant variables as possible, while keeping the number of false positives under control. Well-known examples of multiple-testing correction procedures are the Bonferroni correction to control the family-wise error rate (FWER), and the Benjamini-Hochberg (BH) procedure \cite{benjamini1995controlling} and the Q-value \cite{storey2003statistical} to control the False Discovery Rate (FDR). FDR-controlling procedures are particularly appealing in multivariate settings with large numbers of variables due to their increased statistical power in comparison to FWER-controlling methods. However, applying ANOVA variable by variable not only becomes impractical but also risks missing multivariate patterns that emerge only when multiple variables are considered simultaneously \cite{VisWesterhuisSmilde2007_ASCA_Validation}. 

A popular ANOVA extension for a multivariate-response is ASCA, which combines the design-based decomposition of ANOVA with the dimension{ality} reduction of Principal Component Analysis (PCA). ASCA allows researchers to study factor effects in a way that is multivariate, structured, and interpretable \cite{smilde_anova-simultaneous_2005}. In practice, ASCA makes it possible to answer questions such as: \textit{Which experimental factors significantly affect the overall dataset? How do the groups or treatments separate in the multivariate space? Which variables contribute most to these differences?} The present section provides a systematic workflow that practitioners can follow for understanding how ASCA works, see Figure \ref{fig:DiagPipeline}. 

\subsection{Data curation}
Before applying ASCA to any dataset, it is crucial to perform a thorough check on the data to ensure its quality and integrity. This involves identifying both missing data and potential anomalies (or outliers), as these can significantly impact the results of the factorization and subsequent statistical testing. Note that approaches to deal with missing data and outliers within the ASCA model itself are possible and also discussed in this section.

\subsubsection{Missing data}
Missing data is a challenge in many experimental datasets, and its treatment must be handled with care to avoid introducing bias into the ASCA model. Firstly, it must be noted that null values and below-limit-of-detection values are not missing values; and they must be treated differently, as those numbers have a physical meaning \cite{sun2024pretreating}. Missing values are values that were not recorded, and in this section we refer to missing individual values, not missing samples or variables, which would relate to the unbalanced situation covered in Section \ref{Factorization}. There are several strategies for managing these missing data, each suited for different patterns of missingness and the structure of the data \cite{polushkina-merchanskaya_considerations_2025}. The first approach would be to simply remove the variables or samples that have too many missing values, but this may introduce imbalances in the design, that must be taken into account. 

Another common approach is unconditional mean replacement, where missing values are replaced with the overall mean of the available data. While simple to implement, this method can lead to biased results, particularly if the missing data is not missing completely at random. It is important to note that this method does not account for the underlying structure of the experiment nor for the multivariate nature of the responses, and as such it may introduce unwanted bias.

A more context-sensitive approach is conditional mean replacement. This can be postulated in different ways depending if the condition is made on other responses~\cite{nelson2006impact,arteaga2002dealing} or on the experimental design~\cite{polushkina-merchanskaya_considerations_2025}. For the latter, missing values are imputed using the mean of other observations that share the same factor levels or conditions in the experimental design (that is, observations within the same cell). This method helps maintaining the integrity of the DoE, as it ensures that the imputation is consistent with the existing structure of the data. However, it still does not account for the relationships between responses in multivariate data.

When working with high-dimensional data, multivariate imputation~\cite{nelson2006impact,arteaga2002dealing}, and in particular multiple imputation, may be a more robust method. Multiple imputation generates several different imputed datasets, and the results from these datasets are combined to account for the uncertainty inherent in the missing data. This approach is generally considered more reliable for datasets with significant missingness, as it better reflects the variability in the data. The choice of imputation method should be carefully considered based on the nature of the missing data \cite{NELSON199645}.

Finally, in ASCA implementations based on permutation testing, it is also possible to deal with missing data within the inference step~\cite{polushkina-merchanskaya_considerations_2025}. The basic idea is to use the same imputation approach for true and permuted data, so that the bias generated by the missing data mechanism is considered when creating the null distribution. 

\subsubsection{Anomaly detection}

In addition to handling missing data, it is equally important to detect and manage anomalies in the dataset. Anomalies, such as outliers or other unusual data points, can skew the factorization and undermine the validity of the results. Anomaly detection should be performed both \textit{a priori} (before running ASCA) and {also on the residuals of the ASCA model}, to ensure that the analysis remains robust \cite{VisWesterhuisSmilde2007_ASCA_Validation}.

\textit{A priori} anomaly detection is typically done through exploratory data analysis such as PCA, which helps to visualize the structure of the data and identify observations that fall outside the expected range of variation. By examining PCA score plots and residuals, it becomes possible to spot extreme data points that might not belong to the expected data distribution. The identification of anomalies should be done systematically and having a focus on the context. If outliers result from experimental errors, they should be removed from the dataset. However, if they represent true biological or chemical variation, they should be retained, as excluding them could lead to distorting the reality and the loss of valuable information.

Complementary, the anomaly detection can also be performed on the residuals of the ASCA model. This will also allow to determine if the data should be transformed, as discussed afterwards.

\subsection{Data preprocessing}
\label{preprocessing}
Data pretreatment is a very relevant step in the ASCA pipeline, because it directly affects how variance is distributed in the dataset, and therefore how factor effects will appear in the result. Data preprocessing (if necessary) can have several steps which are commonly applied in the following order: normalization (row-wise preprocessing), transformations (element-wise preprocessing) and scaling (column-wise preprocessing).

Normalization (or row-wise preprocessing) typically addresses systematic effects or artifacts within individual observations. Operations such as baseline correction, smoothing, or alignment, common in spectroscopic and chromatographic data, aim to make samples more comparable, removing non-informative variation unrelated to the experimental design \cite{Rinnan2009TrAC, Cook2014JChemom}. In some omics datasets, dilution effects (e.g., in urine or saliva samples) or an unequal total number of reads (e.g., in metagenomics), make samples incomparable and normalization is necessary~\cite{dieterle2006probabilistic,robinson2010edger}. However, these corrections inevitably modify the variance structure of the data, potentially redistributing variance across model terms and residuals. Therefore, such steps should be applied deliberately, ensuring that only unwanted signal components are removed and that the remaining variance reflects meaningful experimental variation.

Transformations of the signal are considered when the distributional properties of the data strongly deviate from the assumptions underlying ANOVA or ASCA. Most ASCA implementations are based on permutation testing, which as we already describe makes assumptions less stringent. Severe skewness, heavy tails, or heterogeneity of variance across groups can compromise the reliability of ASCA inference \cite{polushkina-merchanskaya_considerations_2025}. To mitigate these effects, transformations such as the Box–Cox power transformation, which stabilizes variance, or the rank transformation, which replaces raw values by their order, may be applied. The rank transformation is especially robust to outliers, while the Box–Cox transformation tends to bring non-normal responses closer to Gaussian behavior \cite{Hoaglin_Mosteller_Tukey_1991,Sakia1992BoxCox}.  A combined form of row-wise preprocessing and transformation is block-wise normalization, which can be applied to different levels in a factor to improve homoscedasticity. Although permutation-based inference makes ASCA relatively tolerant to moderate deviations from distributional assumptions \cite{VisWesterhuisSmilde2007_ASCA_Validation}, appropriate transformations remain a valuable tool in challenging datasets. Their use should be guided by exploratory analysis and diagnostics of data and residual distributions. In some situations, each response should be explored individually to decide whether to use a transformation or not, but most commonly transformations can be applied to groups of similar responses (e.g., spectral responses from the same analytical device) and other approaches to speed up exploration are also possible~\cite{polushkina-merchanskaya_considerations_2025}.

Scaling (or column-wise preprocessing) focuses on standardization of variables to ensure balanced contributions to the model. Mean-centering removes the overall average effect, emphasizing deviations that correspond to experimental contrasts, which aligns with the logic of ANOVA decomposition. However, this step can be performed as part of the ASCA factorization itself just like in ANOVA (see Equation \ref{eq:Anova_regression3}). Scaling, in turn, adjusts for differences in measurement range or variance among variables: it can amplify variables measured on smaller scales or reduce the influence of those with inherently larger magnitudes. Instead of using the overall standard deviation, some alternatives propose using the standard deviation of a reference group or timepoint \cite{timmerman2015scaling}.  While scaling does not affect the standard ANOVA partitioning (with the effects model or {OLS}, as explained in the next section), it does alter the subsequent steps.

Overall, data preprocessing should not be regarded as a routine operation but as a deliberate analytical decision. Each transformation or scaling choice reshapes the variance landscape that ASCA partitions, and careful selection is essential to ensure that the model captures true experimental effects rather than artifacts of preprocessing. {It must also be noted that preprocessing may affect outliers, so data curation is a step that should be present throughout the whole process.}

\subsection{Factorization}

The central step in ASCA is the factorization of the data matrix according to the experimental design. Factorization means partitioning the total variation in the multivariate data into contributions from each factor in the design and from their interactions, plus residual variation. In essence, it is the multivariate extension of what classical ANOVA does with sums of squares in the univariate case \cite{smilde_anova-simultaneous_2005, Hoaglin_Mosteller_Tukey_1991, wilks_MANOVA_1932}. Factorization clarifies the structure: it decomposes the original matrix into an effect matrix for each factor or interaction in an ANOVA fashion:
\begin{equation}
\mathbf{Y} = \mathbf{1} {\mathbf{m}}^T + \mathbf{Y}_A + \mathbf{Y}_B + \mathbf{Y}_{AB} + \mathbf{E}
\label{eq:asca_factorization}
\end{equation}
\noindent where $\mathbf{Y}$ is the original data matrix, {$\mathbf{1} \mathbf{m}^T$ contains the column averages or intercepts}, $\mathbf{Y}_A$ and $\mathbf{Y}_B$ are the effect matrices for {crossed} Factors $A$ and $B$, respectively, $\mathbf{Y}_{AB}$ is their interaction, and $\mathbf{E}$ is the residual matrix. This factorization can be carried out in different ways depending on the design and coding choices. In the classical ASCA framework it is based on ANOVA effect matrices (by averaging), while in ASCA+ the GLM formulation is used, which is a better approach to handle unbalanced designs. If desired, maximum-likelihood extensions can be used to account for random effects (see Section \ref{sub:LinearMixedModels}) \cite{smilde_anova-simultaneous_2005, thiel_asca_2017, martin_limmpca_2020} at the expense of a high computational load. Regardless of the algorithm, the goal remains the same: to partition the total variation into effect matrices that correspond to the experimental factors and interactions of interest, ready for the subsequent analysis and visualization.

In the factorization step, special attention must be put on unbalanced designs. Classical ASCA, like ANOVA, assumes balance, where each condition contributes equally to the estimation of main effects and interactions. When this assumption is violated, effect estimates may become biased, and the orthogonality between factor and interaction matrices is lost. Consequently, part of the variance attributed to one factor may in fact originate from another, leading to misleading interpretations.

Practitioners should therefore carefully inspect the design matrix before model fitting. If imbalance is detected, classical mean-difference decomposition should be avoided {since they generate biased estimations}. Several solutions have been reported to handle {data from unbalance designs}, including GLM formulations such as ASCA+, Weighted Effect ASCA (WE-ASCA) \cite{Ali2020WEASCA} and rebalanced ASCA~\cite{de2023efficiently}. Furthermore, if GLM is used, any form of sequential factorization and SS computation can be considered, either computed directly from the regression {involving all the model terms} or alternative Type I, Type II or Type III computations.

In practice, mild unbalance may not drastically affect qualitative conclusions, but as imbalance increases, factor confounding becomes unavoidable. Always verify group sizes, prefer GLM-based approaches, and interpret explained variance cautiously, acknowledging that part of the variability may be redistributed across effects when the design is not fully balanced.


\subsubsection{Checking the assumptions}

While parametric ANOVA assumes that residuals follow a normal distribution, in ASCA this assumption is less restrictive. This is because the statistical inference is commonly assessed using permutation testing, which is a distribution-free alternative. Therefore, even when residuals deviate from normality, inference remains valid as long as the permutations are based on the \textbf{exchangeability assumption}, which comes down to the independent and identically distributed errors assumption \cite{anderson2003permutation}.

\textbf{Independence of residuals}. Residuals must be independent across experimental units; otherwise, correlations (for example, between repeated measures or nested samples) may inflate the apparent significance of effects. Independence can only be ensured through proper design, not through posthoc corrections, but it also depends on selecting the correct model structure that includes all relevant effects. {A useful way to check this assumption is to plot the residuals against the sample order}, for instance, which may highlight temporal drift or run-order artifacts.

\textbf{Homoscedasticity (equal variance) of residuals}. As in ANOVA, ASCA assumes that the residual variance is comparable across groups. The residuals in the high-dimensional case are usually summarized in the Q-statistic (the sum of squared residuals over all variables). Violations of the homoscedasticity assumption can be examined by plotting boxplots of residuals grouped by factor levels, or by inspecting the residual variance matrix for heterogeneity. Strong differences in residual distributions or outlying samples may indicate that the magnitude of their modeled effects is biased. These problems need to be solved in the data curation or preprocessing steps before interpreting the ASCA table{, usually this can be fixed by transforming the variables adequately}. 
    
These assumptions do not invalidate ASCA when mildly violated, but they do influence the interpretation of significance and variance partitioning. Proper diagnostics and thoughtful design remain the best safeguards for robust inference.

\subsubsection{Evaluating the model}

Once the ASCA model is calculated (and checked that {it} is adequate), and the effect-matrices are available, the effect size of each factor and interaction can be quantified through the SS of its effect matrix. This is, the summation of each individual element in the matrix squared. The rest is computed like in the univariate case, including the percentage of the total SS, which provides a measure of Explained Variance (EV, Equation \ref{eq:expVariance}), DoFs, MS and F-ratios.
\begin{equation}
\text{Explained Variance}_{A} (\%)= 
\frac{\lVert \mathbf{Y}_{A} \rVert^{2}}
     {\lVert \mathbf{Y} - \mathbf{1} {\mathbf{m}}^T \rVert^{2}}
\times 100\% = 
\frac{SS_A}{SS_{\text{tot}}}
\times 100\%
\label{eq:expVariance}
\end{equation}
where $SS_A = \lVert \mathbf{Y}_{A} \rVert^{2}$ is the sum of squares{, or the Frobenius norm,} of the effect matrix for Factor $A$, and $SS_{\text{tot}} = \lVert \mathbf{Y} - \mathbf{1} {\mathbf{m}}^T \rVert^{2}$ is the total sum of squares of the original data matrix after subtracting the mean, which is analogous to the concept of variance. By comparing the proportion of variance explained across different factors, it can be identified which factors have the strongest influence on the multivariate response and which effects are relatively minor. This comparison provides a sense of the relative size of factor contributions, allowing researchers to prioritize the most relevant sources of variation for interpretation. Also the non-explained variance is of high interest, this is, the variance left in the residual matrix $\mathbf{E}$, which illustrates how adequate the applied model is \cite{smilde_anova-simultaneous_2005, VisWesterhuisSmilde2007_ASCA_Validation}.

\subsection{Inference}

The aim of inference testing is to validate that the detected effects are statistically reliable and not artifacts of random variation. While classical parametric ANOVA inference relies on the \textit{F} distribution, in the high-dimensional collinear data in the ASCA situation the assumptions of multivariate normality and normal distribution of residuals are not guaranteed, and parametric tests (where a certain distribution is assumed) may not be always appropriate. Instead, significance is commonly assessed using permutation testing, which builds an empirical null distribution for each effect by randomly permuting the sample labels according to the experimental design (Figure \ref{fig:Fdistribution}). This allows to estimate \textit{p}-values without assuming specific data distributions, defined as the fraction of calculated statistics that fall above the observed value \cite{anderson2003permutation, camacho_permutation_2023}. 

\begin{figure}
    \centering
    \includegraphics[width=0.5\linewidth]{Fdistribution2.png}
    \caption{Illustration of a histogram of statistics obtained under the null hypothesis by permuting group labels. The red vertical line marks the observed statistic for the original model.}
    \label{fig:Fdistribution}
\end{figure}

This statistic can be the EV, SS or the MS, like in univariate ANOVA. However, a more formal statistical framework involves computing an \textit{F}-ratio, which compares the variance explained by an effect to the residual variance. This again links back to univariate ANOVA, where the \textit{F}-ratio is the standard tool for significance testing. As already pointed out, {t}he \textit{F}-ratio for each term is the ratio between the MS of that term and a reference MS. The selection of the reference  depends on the fixed/random nature of the terms and the relationships among terms~\cite{anderson2003permutation}, like we illustrate in the example of BOX 3. In ASCA (multivariate data), an analogous statistic can be derived to assess whether the variation captured by an effect is larger than would be expected by chance (Equation \ref{eq:fStatistic}) \cite{VisWesterhuisSmilde2007_ASCA_Validation, camacho_population_2024}. 

\begin{equation}
F_A = \frac{SS_A / dof_A}{SS_R / dof_R}
\label{eq:fStatistic}
\end{equation}

\noindent where $SS_A$ is the sum of squares for Factor $A$, $dof_A$ its DoFs, $SS_R$ the reference sum of squares, and $dof_R$ the reference DoFs. When all factors are fixed, the reference of all model terms are the residuals. Random factors in the design make the selection of the reference more complex~\cite{montgomery2020design}. 

Once the statistic is selected, another relevant element in the inference is the permutation strategy. Four main types of permutation strategies have been described for ASCA and related models~\cite{rasmussen2024_permutation_asca,anderson2003permutation}:

\begin{enumerate}
    \item Unconstrained permutations, in which row-wise permutations are performed completely at random either over the data matrix $\mathbf{Y}$ or over the {model (or coding)} matrix $\mathbf{X}$, leaving the other fixed.
    \item Constrained permutations, in which row-wise permutations are only allowed among exchangeable units under the null hypothesis.
    \item Residual permutations, in which row-wise permutations are performed on the residuals of a factorization that does not include the term been tested, quite often the reduced model, that is, a factorization with all the other terms.
    \item Marginalized permutations, based on a reduce{d} model factorization that orthogonalizes the coding matrices of the different terms.
\end{enumerate}

{From the permutation results, the p-value is then calculated as:
\begin{equation} \label{eq:pvalue}
p_A = \frac{\#\{F_A^k \geq F_A; \; k=1,\dots,K\} + 1}{K + 1},
\end{equation}
where $F_A$ is the F-ratio for Factor $A$, $F_A^k$ denotes the same F-ratio obtained in the $k$-th permutation, and $K$ is the total number of permutations performed. The p-value provides an empirical estimate of the probability of observing a result as extreme or more extreme than the one obtained, assuming the null hypothesis is true.}

It should be noted that different software packages make use of different permutation approaches~\cite{maddahi5362895investigation}, which may lead to confusion and lack of reproducibility in the ASCA literature. Unconstrained permutations are the fastest and simple{st} to program, because it takes a single permutation run to obtain permuted statistics for all model terms and residuals. All other permutation policies make use of different factorizations to test different terms, which means that they require $T$ times more permutations than unconstrained permutations, with $T$ the number of terms in the model. {In particular, general solutions for any design are complex to program for constrained permutations, and to a lesser extent, marginalized permutations.} 

Unconstrained permutations, however, do not comply with exchangeability constrains while all the other methods do. As a result, they only provide approximate tests even for factors for which exact tests exist. However, it should be noted that an exact test is not always the most powerful solution, since the number of possible permutations that fulfill exchangeability may be too low. Indeed, statistical power for a controlled Type I error is probably the most important feature in an inference system. In numerical practice, residual permutations under the reduced model have been reported to provide very powerful estimates in balanced ANOVA~\cite{anderson2003permutation}, while in unbalanced ASCA they have been shown to be less reliable~\cite{camacho_permutation_2023}. In comparison, unconstrained permutations yield statistically powerful estimates in both balanced~\cite{maddahi5362895investigation,camacho_population_2024} and unbalanced~\cite{camacho_permutation_2023} ASCA.

\subsection{Visualization and post-hoc testing}

Visualization of the factor effect is a key strength of ASCA. Once the total data matrix has been factorized according to the experimental design, in a multivariate-ANOVA fashion (Equation \ref{eq:asca_factorization}), each significant effect matrix ({e.g.,} $\mathbf{Y}_A$) is modeled separately using Simultaneous Component Analysis (SCA). SCA is mathematically analogous to Principal Component Analysis (PCA), but it is applied 
to each term effect or combinations thereof, including the measurements corresponding to all levels/cells in the terms (thus the name "simultaneous"). This means that the dimensionality reduction and visualization of each factor are performed after the ANOVA decomposition, not as part of it. Consequently, the visualization does not alter the ANOVA structure or the partitioning of variance; it merely provides a compact, interpretable summary of the multivariate patterns captured by each factor \cite{smilde2005multi}. Each SCA model extracts a small number of components summarizing the variation explained by the corresponding factor following Equation \ref{eq:SCA}:
\begin{equation}
\mathbf{Y}_A = \mathbf{T}_A \mathbf{P}_A^{\mathrm{T}} + \mathbf{E}_A
\label{eq:SCA}
\end{equation}
where $\mathbf{T}_A$ is the scores matrix for Factor $A$, $\mathbf{P}_A$ is the loading matrix and $\mathbf{E}_A$ is the residual matrix. Score plots show how the group means are distributed according to that effect, while loading plots indicate which variables (e.g., wavelengths, metabolites, or sensors) are most responsible for the observed differences. Together, they allow practitioners to visualize how a given factor contributes to the structure of the data. The maximum number of components in model (\ref{eq:SCA}) is the number of DoFs {o}f the term, which comes from the constrains imposed in the ANOVA coding. 

However, it is important to recognize that the SCA models obtained from each effect matrix do not directly represent uncertainty in the same way as PCA models based on raw data. Each effect matrix in ASCA ({e.g., }$\mathbf{Y}_A$) contains group-mean profiles computed during the ANOVA decomposition. As a result, when SCA is performed on these matrices, the model reflects the average multivariate structure associated with each factor, but not the variability among individual replicates in a group. In other words, the raw SCA model only summarizes systematic effects, with no uncertainty between the replicates visualized. 
An interesting development of ASCA plots is to augment scores with residual variation ($\mathbf{T}_A^a = (\mathbf{Y}_A + \mathbf{E}) \mathbf{P}_A$, where $\mathbf{T}_A^a$ represents these augmented scores) to provide a baseline to assess when the separability among levels or cells is significant. For instance, well-separated groups in an augmented score plot indicate that the factor clearly differentiates those conditions, while overlapping groups suggest that these levels are not well separated \cite{thiel_asca_2017, VisWesterhuisSmilde2007_ASCA_Validation}. Thus, augmented score plots are akin to how post-hoc tests work, and ASCA post-hoc tests have been proposed~\cite{ liland2018_asca_conf_ellipsoids}. It should be noted that ASCA scores do not necessarily have to be augmented with residuals, and the most sensible practice is to augment them with the suitable reference term in the denominator of the F-ratio (Equation \ref{eq:fStatistic}). {This is relevant for models that include random factors}. We illustrate this idea in the example of BOX 3.

When interpreting SCA plots:
\begin{itemize}
    \item \textbf{Augmented Scores $\mathbf{T}_A^a$} describe patterns among levels (between-groups variability) for a specific Factor (\textit{A} in this case) compared to the within-group variability (reference in the denominator of the \textit{F}-ratio ). 
    \item \textbf{Loadings $\mathbf{P}_A$} indicate which variables contribute most to the between-groups variability in the score plot. Large positive or negative loadings reveal features (e.g., spectral regions, compounds) driving group differentiation.
    \item \textbf{Residuals $\mathbf{E}_A$} represent unmodeled variation of the effect matrix $\mathbf{Y}_A$. In cases with a large number of levels for the factor, not all principal components will be used in the SCA model of the effect matrix. In those cases a Multivariate Statistical Process Control (MSPC) plot~\cite{ferrer2007multivariate} may be built to visualize the factorization showing the $D$ and $Q$ statistics (for the model and the residual part) {
    \begin{equation}
        D_i = (\boldsymbol \tau^a_i)^{T} (\mathbf{T}_A^{T}\mathbf{T}_A)^{-1} \boldsymbol \tau^a_i 
    \end{equation}}
    {
    \begin{equation}
        Q_i = \boldsymbol (e^a_i)^{T} \boldsymbol e^a_i 
    \end{equation}}
    
    where $D_i$ and $Q_i$ are the $D$ and $Q$ statistics for the $i$-th experimental unit, respectively, $\mathbf{T}_A = \mathbf{Y}_A\mathbf{P}_A$ represents the scores of the PCA model of the factor matrix, and $(\boldsymbol \tau^a_i)^{T}$ and $\boldsymbol e^a_i$ represent the augmented score and residual vectors for the $i$-th experimental unit, respectively. These statistics can indicate outlying samples for this factor. Large residual variance implies that the model only partially explains the observed structure; maybe due to large natural variability, specially in a factor with high DoFs (like is often the case for random factors).

 \item \textbf{Post-hoc tests of significant factors}
    Similar to ANOVA, after a factor has been found significant, post-hoc tests can be used to test which of the levels are statistically significantly different. If the ASCA model of the significant factor uses a single principal component, similar methods as used in ANOVA apply. In case multiple principal components are used to describe the variation in the effect matrix of that significant factor, confidence ellipses around group means can be used \cite{liland2018_asca_conf_ellipsoids}. Other approaches combining multivariate and univariate statistics are also possible~\cite{mahieu2025_multanova}.
    
    \item \textbf{PCA on the residual matrix of ASCA}
    A PCA analysis of the residual matrix of the ASCA model ($\mathbf{E}$) can provide information on outlying behavior, significant effects in the residuals{, or the violation of the assumptions described in Section 4.3.1}. Such effects can occur when multiple variables deviate in the same way from their group average. An example is that replicate plants are not perfectly of equal size and thus levels of multiple metabolites might all be higher in the larger plant while they are all smaller in a smaller plant{, causing correlation between various metabolites}. Such an effect was shown in the Caldana paper for phenylalanine and shikimic acid under certain conditions \cite{caldana2011high}. Furthermore, such correlation between those metabolites might be different in different cells of the design, as was shown in the same paper. Note that these correlations do not violate the sample independence assumption. They appear due to correlations of different variables, an effect not modeled in the factors of the design. 
\end{itemize}

 \begin{tcolorbox}[breakable,title={BOX 3: Guiding example: ASCA analysis},colframe=green!50!black]

Let us come back to the ASCA analysis of the example in BOX 2. The example is performed with the MEDA Toolbox v1.11~\cite{MEDARep,camacho2015multivariate}, a software package for Matlab that is openly available to the community. {Please, note that there are several software packages that include ASCA models with relevant differences in model estimation (averages, regression or maximum likelihood) and inferential procedures (mostly, different permutation techniques). A list of such packages can be found the recent book of Smilde et al.~\cite{smilde2025analysis}. The MEDA Toolbox v1.10 makes use of regression estimates (based on {OLS}) and inferential procedures based on unrestricted permutation of the raw data.}


\begin{itemize}

\item \textbf{Data curation}: An initial exploration for outliers in the original data is shown in Figure \ref{fig:outliers}(a). While we could discard some potential outliers, we double-check this result in the residuals of the ASCA model.

\item \textbf{Preprocessing}: No normalization (row-wise preprocessing) was applied to the data. Data was auto-scaled to homogenize the load of each metabolite in the analysis. 

\item \textbf{Factorization}: Factorization was performed according to the following model:   
\begin{equation} \label{eq:window}
\bY  =  \mathbf{1}{\mathbf{m}}^T + \mathbf{Y}_{Responder} + \mathbf{Y}_{Time} + \mathbf{Y}_{RT} + \mathbf{Y}_{Patient(Responder)}+ \bE
\end{equation}
In this model, we assume the Factor Patient is nested in Factor Responder, and we include the interaction between Responder and Time $\mathbf{Y}_{RT}$. The model corresponds to a GLM factorization of a repeated measures model. 

\item \textbf{Checking assumptions}: Figure 1 in Supplementary Materials illustrates different plots to check the assumptions in the residuals. In particular, the residuals exploration in terms of the patients is shown in Figure \ref{fig:outliers}(b). Patients M0370, M0357 and {M0291}, also spotted before as potential outliers, show very large residuals in one of their time points. We tried both (i) to discard {all samples from these outliers, with a remaining sample of 11 responders and 7 nonresponders}, and repeat the factorization and (ii) to apply the rank transformation to the complete data, as suggested by Polushkina et al.~\cite{polushkina-merchanskaya_considerations_2025}. We found very similar results in terms of factorization and the assumptions on the residuals, {which validates the result of both analyses}, so we opted for the first choice in order to maintain better interpretability of the data. 

\item \textbf{Inference}: After factorization and inference, we obtain the ASCA table (Table \ref{tab:ASCA3}). The SS reported is the one obtained for the simultaneous factorization of all the model terms, without applying any correction (like Type I, II or III SS). We can see that the total $\%SS$ takes slightly more than the expected value of $100\%$, which comes from the imbalance in the data. This reflects double-counting of variance, small enough so that no correction needs to be applied. In the example, the F-ratio for all terms except for the Factor Responder is computed using the MS of residuals as the reference (Equation { \ref{eq:fStatistic}}). In the case of Factor Responder, however, the reference is Factor Patient, since it represents  the only random term below Responder in the Hasse diagram (Figure \ref{fig:Hasse})~\cite{anderson2003permutation, montgomery2020design}. Both Factors Responder and Patient are significant. This is generally expected for Factor Patient, due to the commonly high individual variability in clinical studies. The significance of Factor Responder is more interesting, since it means that there are some biomarkers (metabolites) that distinguish responders from non-responders in a way that is not expected to be the result of randomness.   

\item \textbf{Visualization}: We proceed with the visualization of the terms that were found significant. Note that given the supervised nature of ASCA, non-significant terms should not be visualized. We start with the visualization of Factor Responder in Figure \ref{fig:Responder}. {The PCA {components} are calculated} from the data of $\bY_{Responder}$ but the scores are augmented as $\bY_{Responder}+\bY_{Patient}$, since Factor Patient is the reference in the F-ratio. Since the Factor Responder has only one degree of freedom, we can only fit a model with one component. The score plot clearly shows the separation between responders (towards the negative direction) and non-responders (towards the positive direction). The loadings are interpreted accordingly: metabolites towards the positive direction are generally more abundant in the non-responders, and the other way round. We continue with the visualization of Factor Patient in Figure \ref{fig:Patient}. In this case, the model is built from $\bY_{Patient}$ but the scores are augmented as {$\bY_{Patient}+\bE$}. The model can have as many PCs as the DoFs of the factor, 16 in this case. Yet, looking at the scree plot we decided to visualize the first 2 PCs in the score and loading plot (Figures \ref{fig:Patient}(a) and \ref{fig:Patient}(b)) and then use a MSPC plot to visualize model D-st vs residual Q-st (Figure \ref{fig:Patient}(c)). Some patient-specific patterns can be inferred from the plots, like for patient M0271, {for which all measurements  have a extreme negative score in PC1, which could be associated to specific metabolites (e.g., high concentration of the metabolite at 98.9772)}.  

\end{itemize}

\end{tcolorbox}

\begin{figure*}[h]
 	\centering
 	\subfigure[PCA]{\includegraphics[width=0.45\textwidth]{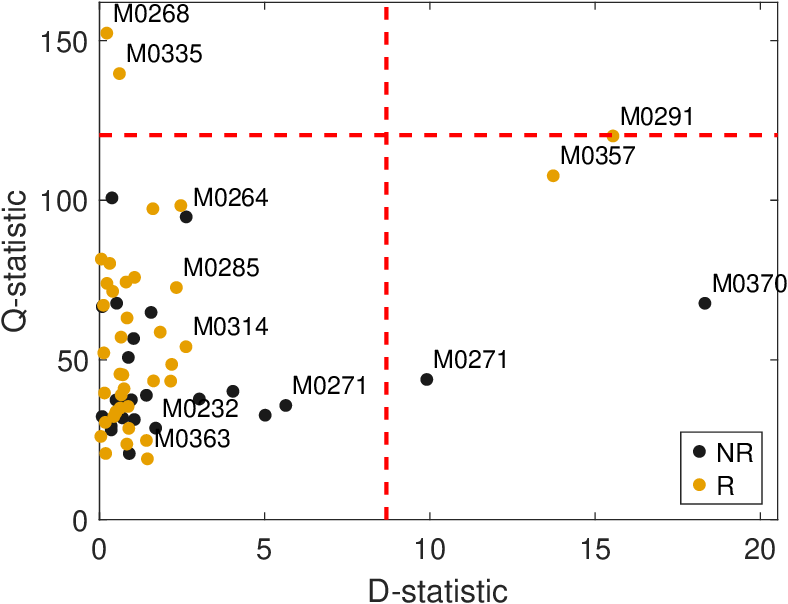}}
 	\subfigure[ASCA]{\includegraphics[width=0.45\textwidth]{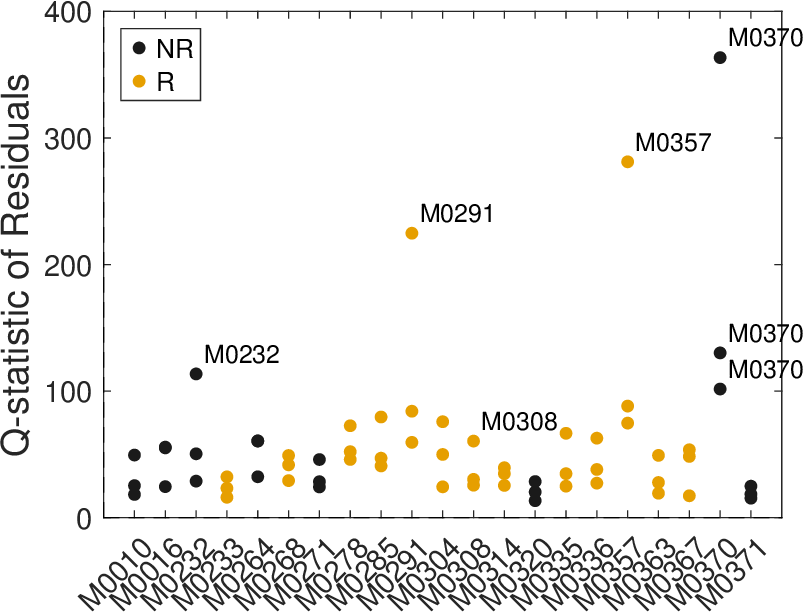}}
     	\caption{Exploration for outlier detection with (a) PCA-based Multivariate Statistical Process Control (MSPC) chart of the data {matrix $\mathbf{Y}$ and (b) Exploration of ASCA-based residuals $\mathbf{E}$ using the Q-statistic and colored in terms of the Patient. NR stands for non-responders and R for responders}.}
\label{fig:outliers}
 \end{figure*}

\begin{table}[h]
    \caption{ASCA table of the data without three outliers. The table includes the statistical decomposition and inference information, following a traditional format that includes: the sum of squares (SS) associated with each term (factors and interactions), its percentage of the total sum of squares (\%SS), the number of degrees of freedom ({DoFs}), the mean sum of squares (MSS), the F-ratio (F) and the p-value. {NR stands for non-responders and R for responders.}}
     \label{tab:ASCA3}
    \centering 
    \begin{tabular}{lrrrrrrr}
 & SS & $\%SS$ & DoFs & MS & F & p-value \\ 
 \hline 
 \hline 
Factor Responder & 451 & 7.6 & 1 & 451 & 2.8 & 0.01 \\ 
Factor Time & 150 & 2.5 & 2 & 75 & 0.9 & 0.56 \\ 
Factor Patient & 2579 & 43.5 & 16 & 161 & 2.0 & $< 0.01$ \\ 
Interaction Responder $\times$ Time & 205 & 3.5 & 2 & 102 & 1.3 & 0.26 \\ 
Residuals & 2609 & 44.0 & 32 & 82 &  &  \\ 
 \hline 
Total & 5936 & 101.1 & 53 & 110 &  &  \\ 
 \hline 
 \hline 
    \end{tabular} 
\end{table}

   \begin{figure*}
 	\centering
	\subfigure[Scores]{\includegraphics[width=0.45\textwidth]{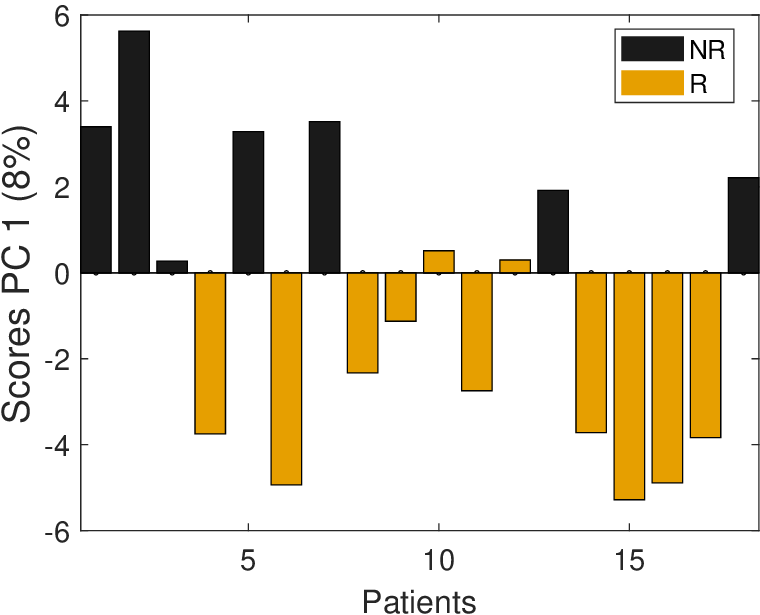}}
	\subfigure[Loadings]{\includegraphics[width=0.45\textwidth]{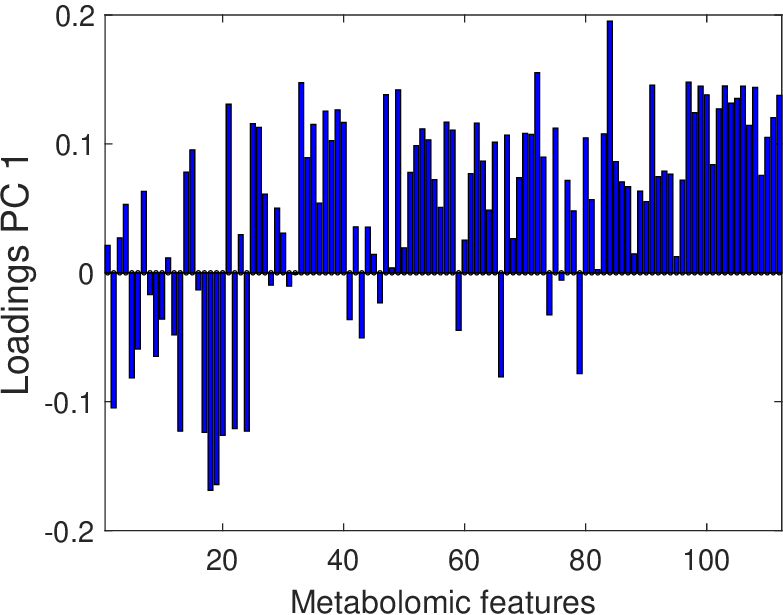}}
 	\caption{Factor Responder: ASCA score plot (a) and loading plot (b). {NR stands for non-responders and R for responders}.}
\label{fig:Responder}
 \end{figure*}

   \begin{figure*}
 	\centering
	\subfigure[Scores]{\includegraphics[width=0.45\textwidth]{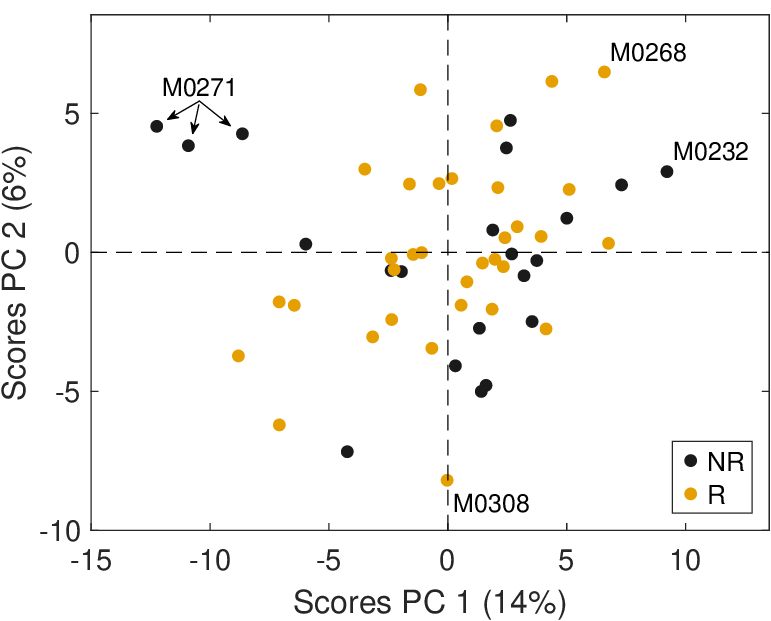}}
	\subfigure[Loadings]{\includegraphics[width=0.45\textwidth]{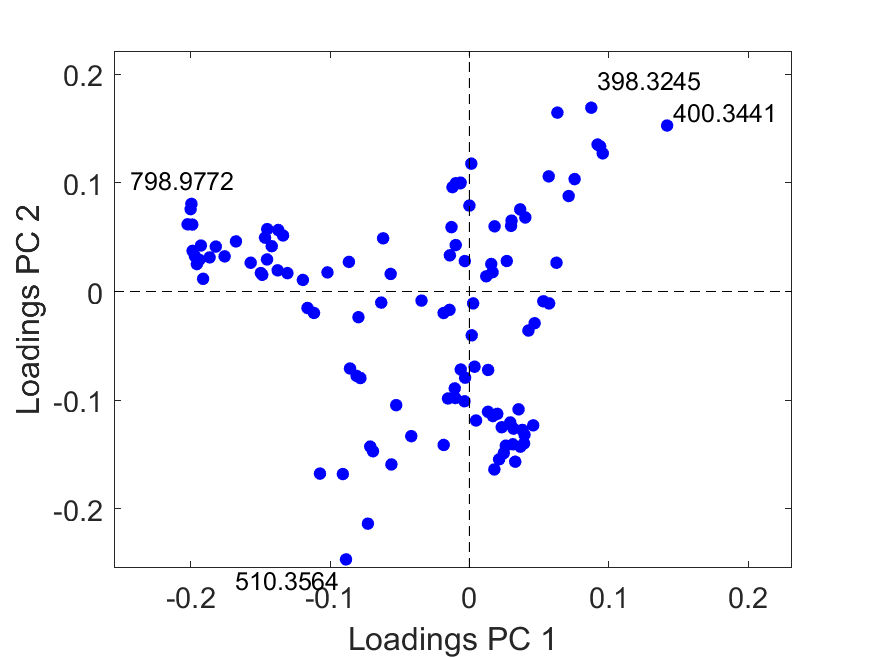}}
	\subfigure[MSPC]{\includegraphics[width=0.45\textwidth]{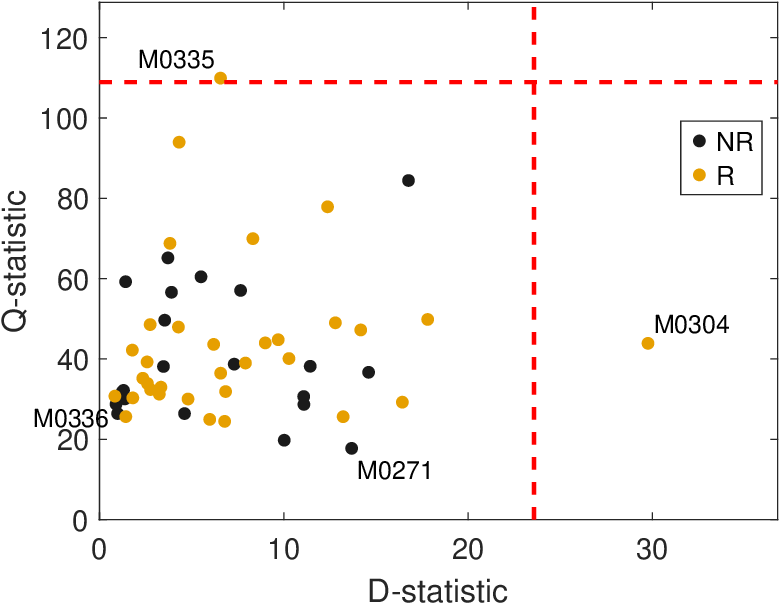}}
 	\caption{Factor Patient: ASCA score plot (a), loading plot (b) and MSPC plot (c). {NR stands for non-responders and R for responders}.}
\label{fig:Patient}
 \end{figure*}

\section{Alternative methods and extensions} \label{Related}

There are many alternative methods for the analysis of variance for high dimensional data. Most of these methods use alternative ANOVA models, have differences in the PCA analysis of the effect matrices, transform the multivariate data to between sample distances or to ranks. For some methods many parameters have to be set, while others are easy to use. The overview of methods provided is certainly not complete but aims to group some of these methods in how they deviate from ASCA.

\subsection{Methods based on distances}
There are multiple alternatives for ASCA, to model multivariate data from a predefined DoE with multiple factors. The first set of approaches, specifically popular in ecology and meta-genomics are ANOSIM \cite{clarke_non-parametric_1993} and PERMANOVA \cite{anderson_permanova_2013}, which reduce the high dimensional data into a sample by sample distance matrix. PERMANOVA works by comparing distances between samples of different groups to distances between samples of the same group. The test compares these distances. The inference is performed using permutation testing. While powerful, working on the distances limits our ability to relate objects with variables. ANOSIM is a non-parametric test based on ranks of the distances and is relatively inflexible for complex designs.

\subsection{MANOVA type methods}
For the situation where the number of variables is limited, MANOVA can be used. For one-way {M}ANOVA, the "within group" covariance matrix and the "between group" covariance matrix are obtained from the data. The test depends on whether the between group variance is large enough compared to the within group variance, similar to an F-test in ANOVA. The within group covariance matrix has to be invertable to estimate the test, and therefore the number of variables has to be smaller than the number of samples. An extension of MANOVA called Regularized MANOVA (rMANOVA)~\cite{engel2015regularized} mixes the calculated within group covariance matrix with a target covariance matrix, to make their sum invertable again, and thus applicable to high dimensional data.

\subsection{Methods with altered way of inference}
{ANOVA}-Principal Component Analysis (APCA) \cite{harrington2005analysis} decomposes the response data matrix $\mathbf{Y}$ step by step similar to ASCA using the ANOVA model, considering orthogonality between the factors. The PCA step, however, is applied to {the sum of the effect matrix and the Pure Error matrix (which equals the residual matrix in ASCA), which changes the visualization. The scores of this matrix directly contain the between-replicate variations. When multiple effect matrices are visualized by PCA, their results are not independent due to the addition of the same Pure Error matrix. The power of APCA, and the retrieval of underlying factors was shown to be inferior to that of ASCA \cite{zwanenburg2011anova}}.

{The inference is then applied in a way clearly described by Climaco-Pinto et al. \cite{climaco2009improving}. They also show an improvement of the APCA approach by selective elimination of systematic structures in the Pure Error matrix to get a better estimate of the Pure Random Error \cite{climaco2009improving}}. REP-ASCA uses a similar approach to estimate the systematic variation components in the Pure Error from a second data set with many measurements from identical samples mimicking expected uncontrolled variation \cite{ryckewaert2020reduction}.

\subsection{Methods with different ANOVA models and codings}
Other alternatives for ASCA use alternative coding schemes in case baseline treatments or baseline timepoints are present in the design. 
Principal Response Curves (PRC) \cite{van1999principal} and Scaled to Maximum and Aligned Reduced Trajectories (SMART) \cite{keun2004geometric} use an ANOVA model with a reference coding approach. Both approaches are applied to designs combining different treatment levels and monitored over multiple time points. In PRC the whole time profile of the Control treatment is forced to 0, while the other treatments are shown to deviate from the control. SMART forces all treatments to be equal at the first time point. Although the factorization of both models and also ASCA differ, they end up with the same explained variations and residuals. When only a reduced set of principal components of the effect matrices are visualized, the methods give slightly different results.

Multilevel Simultaneous Component Analysis is an ASCA model based on a nested design where Factor $B$ representing the “within variation” is nested in Factor $A$ (representing the “between variation”) \cite{jansen2005multilevel}. Such a model lacks an interaction.

Partial Redundancy Analysis is an alternative approach to calculate the ASCA model especially in non-orthogonal designs due to specific codings or unbalancedness. It is a multivariate regression with a rank-restriction on the matrix of regression coefficients. It performs ASCA+ as a one-step approach where the net contribution of the effect matrix of interest is forced to have a reduced rank \cite{ter_braak_redundancy_2023}. In the non-orthogonal (unbalanced) case in ASCA+, effect matrices can be non-centered. If this is the case, our recommendation is to fit PCA directly on the non-centered matrices, that is, without performing an additional centering step (this is the approach of the example of BOX 3), since the goal of PCA visualizations is to understand inter-level differences and the location of the center of coordinates in not relevant in practice. Furthermore, a centering step would make the visualizations inconsistent with the SS reported in the table. Furthermore, Partial Redundancy Analysis uses a projection step that can introduce new variation not related to the effect of interest, and therefore should be used with care. 

{Partial Redundancy Analysis is a similar framework to ASCA  \cite{ter_braak_redundancy_2023}, more popular in ecology. Partial RDA is typically applied as a sequential process, orthogonalizing the data to remove unwanted variability before factorization. While in principle both frameworks can handle data similarly (ASCA can also be computed sequentially), Partial RDA traditionally emphasizes the removal of unwanted variability, whereas ASCA, as an extension of multi-way ANOVA, excels in decomposing complex experimental designs.}

\subsection{Methods using maximum likelihood in mixed models}
The Repeated Measures ASCA+ (RMASCA+) method applies the Restricted Maximum Likelihood (REML) algorithm to a repeated measures mixed model for each of the variables in the multivariate data set \cite{madssen_statistical_2025}. The fixed effect estimates are then collected in effect matrices. Inference and visualization of fixed factor effects are performed in a similar way as in ASCA+.

Linear Mixed Model Principal Component Analysis (LiMM-PCA) is an alternative approach for ASCA+ when a REML mixed model is used to describe the variation in the data \cite{martin_limmpca_2020}. As applying maximum likelihood mixed models for many variables is costly with respect to time, in LiMM-PCA an initial PCA step is applied to the high-dimensional data to reduce the dimension of the data set. Then, the mixed model is applied to each of the principal component scores of the data. Using the PCA loadings, it is possible to perform statistical inference on the scores level, while interpretation of the effects can be done of the measured variables.

Another difference between the RMASCA+ and LiMM-PCA model is the procedure to show statistical significance of their main factor effects. A recent comparison study \cite{madssen_statistical_2025} showed that the global log likelihood ratio (GLLR) test, originally promoted in the LiMM-PCA paper {along with the use of bootstrapping}, may deteriorate when the number of principal components in the first step becomes too large, while permutation tests {based on the F-ratio} seemed to perform consistently.

\subsection{Methods with different visualization}

The PARAFASCA (PARAFAC-ASCA) approach extends ASCA by combining its ANOVA-based variance decomposition with a Parallel Factor Analysis (PARAFAC) model to capture complex multifactor interactions \cite{Jansen2008_PARAFASCA}. Interactions involving multiple crossed factors can be rearranged into a data tensor which modes are a combination of the factors (e.g., $A$, $B$) and the response variables, allowing PARAFAC to provide a more parsimonious and interpretable trilinear model with fewer parameters. The main advantage of PARAFASCA lies in its simultaneous representation of all involved factors through mode-specific score vectors that can be visualized in parallel component plots or trilinear biplots, facilitating direct interpretation of variable responses across treatments and time \cite{ArmstrongCamacho2024_PARAFASCA_Replicates}. Note, however, that not all interactions need to be trilinear (or multilinear). Unlike PCA, PARAFAC models are rotationally invariant, requiring careful component selection via explained variance, core consistency, or permutation-based validation. Interpretability strongly depends on experimental design quality-balanced replication and consistent scaling across modes are essential for reliable trilinear patterns. A detailed comparison of PARAFASCA with other ASCA-type methods is provided by Guisset et al. \cite{guisset2019comparison}

In longitudinal or repeated-measures designs, the evolution of multivariate responses over time can be effectively visualized using trajectory plots. Within the RMASCA+ framework, plotting the score trajectories of individual subjects or group means across time points provides a direct representation of within-subject dynamics and temporal trends. These trajectories often reveal treatment effects or recovery patterns that are not evident in static score plots.

For high-dimensional omics or metabolomics datasets, sparse or group-wise visualizations, as implemented in GASCA~\cite{saccenti2018group}, have proven particularly advantageous. By enforcing sparsity or grouping correlated variables, these models highlight the most relevant variable subsets, leading to clearer and more interpretable loading structures while reducing visual clutter in standard biplots.


\subsection{Other ASCA extensions}

\subsubsection{MASCARA}
Multivariate ASCA Residual Analysis is a new extension in which ASCA is used as a filter \cite{white2025mascara}. MASCARA can be applied to filter the major variation in the data that is caused by some factors with large effect, to find relevant but subtle variations in the residual matrix. The residual matrix (containing the information of interest) was applied for co-expression analysis of genes of a specific pathway. Some known genes of this pathway are used as "baits" to find other related genes of the same pathway. The latter is obtained by a Partial Least Squares (PLS) model between the bait genes as the response $\mathbf{Y}$ in the PLS model and all other genes as the predictors $\mathbf{X}$. The high loadings of this PLS model point to the co-expressed genes of interest.
 
\subsubsection{Variable Selection (vASCA)}
ASCA applies a "holistic" significance test for each model term, considering all response variables simultaneously. In high-dimensional datasets, such as in omics sciences (genomics, proteomics, metabolomics), many responses might not be affected by a model term, diluting the statistical signal and overlooking effects in only a few responses or potential biomarkers. Variable-selection ASCA (vASCA)~\cite{camacho_variable-selection_nodate} is an extension of ASCA designed to enhance its statistical power in such situations. By filtering out variables that do not contribute to a particular effect, vASCA is shown to be more powerful than both conventional ASCA (thanks to the variable selection) and to multiple univariate testing like FDR-controlling procedures (thanks to vASCA's multivariate nature). vASCA can be seen as a generalization of ASCA where only the most promising subset of responses is used for inference and visualization. Clearly, this subset may be different for different model terms.  

\subsubsection{Unfolding}
Before applying ASCA, data can be handled in many ways, as explained in Section \ref{preprocessing}. Data unfolding is a step that converts a multiway data tensor (e.g., patients × responses × time) into a two-way matrix so ASCA can be applied on it. However, there are several ways of unfolding: the tensor can be unfolded along different modes (dimensions) and in different manners. A sample-wise unfolding (e.g., rows = patients × time, columns = responses) emphasizes how the measured profiles change according to the full design matrix; a variable-wise unfolding (e.g., rows = patients, columns = responses × time) highlights how variables evolve across repeated measures. Unfolding data can be a way of handling nested factors, or a way of focusing on different aspects of the data; if for instance, a factor is so big that its effect masks the rest \cite{Ezenarro2024NectarinesASCA}.
 
\subsubsection{Power Curves}
The statistical power of a test is a central concept in any inferential procedure. As already described, the power of a test measures the probability that it correctly rejects the null hypothesis when the alternative hypothesis is true.  Power curves typically represent statistical power as a function of the sample size—that is, the number of replicates or experimental runs under the same conditions. This analysis is recommended prior to any research experiment to determine the required number of subjects and experimental levels needed for a desired statistical power. In standard univariate ANOVA, or when relatively few responses are considered, it is possible to use analytical methods to derive power curves. However, in the context of multivariate responses, numerical methods \cite{anderson2003permutation} are a viable alternative. ASCA power curves~\cite{camacho_permutation_2023,camacho_population_2024} can be used to visualize statistical power as a function of the effect size of the model terms or of the sample size. Power curves allow us to make decisions during the experimental planning phase regarding critical aspects of a multivariate design, including the number and nature (fixed/random) of factors, their relationships (crossed/nested), the number of levels, and the number of replicates, among others. Power curves are also useful to optimize the inferential engine, including the model terms, the statistics or the permutation strategy.


\section{Conclusions} \label{Conclusions}

This paper offers a tutorial review of ANOVA Simultaneous Component Analysis (ASCA), a methodology recognized as one of the most impactful chemometric tools developed in recent decades. As a tutorial, its primary goal is to establish good practices for ASCA application by firmly grounding them in the foundational theories of Design of Experiments (DoE) and Analysis of Variance (ANOVA). The inspiration ASCA draws from ANOVA has already spurred relevant methodological innovations; by formally connecting ASCA with the powerful DoE/ANOVA tandem, we aim to significantly enhance its utility and cement its role as an indispensable tool in modern experimental science. As a review, we conclude with a discussion of ASCA alternatives, some of them more established tools, and methodological extensions.

\section*{Acknowledgements}

Dr. Pedro Sánchez-Rovira, Dr. Caridad Díaz, and Dr. Carmen González-Olmedo are kindly acknowledged for sharing the data for the leading example, and ROCHE is acknowledged for funding the study. The reviewers are gratefully acknowledged for their useful and detailed reviews. This work was supported by grant no. PID2023-1523010B-IOO (MuSTARD), funded by the Agencia Estatal de Investigación in Spain (call no. MICIU/AEI/10.13039/501100011033), and by the European Regional Development Fund. This work was partially funded by the Novo Nordisk Foundation (grant number NNF21OC0065495). Funding for the open access charge was provided by the Universidad de Granada / CBUA.

\section*{Software}

The leading example of the tutorial can be reproduced using the code at repository \url{https://github.com/josecamachop/ASCATutorial}. This repository requires the MEDA toolbox v1.10 at \url{https://github.com/codaslab/MEDA-Toolbox}. Please, make a reference to \cite{diaz2022predicting} if you make use of the data.

\appendix

\section{Other forms of coding}

\subsection{Reference coding}

Reference coding is often used in the situation where one of the levels of the factor is the control level and the other levels are different treatments. In this case one wants to know the difference between the treatment effect compared to the control effect. The $\alpha$ coefficient for the control level (level 1) is put to 0, and thus only $\mu, \alpha_2$ and $\alpha_3$ parameters have to be calculated. {In that case the model for each group becomes:}

{
\begin{gather}
\mathbf{y}_1 = \mathbf{1}_{n_1}\mathbf{\mu} + \boldsymbol{\epsilon}_1 \\ \nonumber
\mathbf{y}_2 = \mathbf{1}_{n_2}\mathbf{\mu} +\mathbf{1}_{n_2}\mathbf{\alpha}_2  + \boldsymbol{\epsilon}_2 \\ \nonumber
\mathbf{y}_3 = \mathbf{1}_{n_3}\mathbf{\mu} + \mathbf{1}_{n_3}\mathbf{\alpha}_3 + \boldsymbol{\epsilon}_3 \\ \nonumber
\label{eq:Anova_regression4}
\end{gather}}


This leads to the coding matrix in Table \ref{Reference_Coding_matrix}. Again there is no column for $\mathbf{x}_1$, as $\alpha_1$ is defined $0$. This type of coding is called Reference coding.

\begin{table}[h]
\caption{Reference Coding matrix $\mathbf{X}$}
    \label{Reference_Coding_matrix}
    \centering 
    \begin{tabular}{llll}
        \toprule
        \textbf{Group} & $\mu$ & $\mathbf{x}_2$ & $\mathbf{x}_3$ \\
        \midrule
        Group 1 samples & 1 & 0 & 0 \\
        Group 2 samples & 1 & 1 & 0 \\
        Group 3 samples & 1 & 0 & 1 \\
        \bottomrule
    \end{tabular}
 \end{table}

\subsection{Weighted coding}

The weighted coding makes the interaction of Factors $A$ and $B$ orthogonal to the main factors for $A$ and $B$, and also all factors and interactions are orthogonal to the intercept of the model.
To make factors orthogonal to the intercept (which is always coded as a column of ones) it must be mean-centered with mean 0. The values used in the coding matrix for each level of the factor thus has to be balanced such that the sum the column in the coding matrix becomes 0. If Factor $A$ has two levels, and level 1 has 3 replicates and level 2 has 2 replicated, then the coded value for level 1 becomes -2/3 for each replicate and 1 for each replicate of level 2, as 3×-2/3 + 2×1 = 0.

\section{Type I, II  and III SSs}

To properly explain these different approaches for SS computation, we will resort to the following notation and the example in Figure \ref{fig:Hasse}: Let $E(.)$ represent the residual SS for a model. Thus, $E(A, B, C(A), AB)$ is the residual SS fitting the whole model, $E(A)$ is the residual SS fitting the main effect of A only, and so on. In Type I SS, we apply a sequential order in the factorization and the effect size estimates, so that the SS of each term is calculated from the residuals after the SS of previous terms in the order has already been removed. The literature disagrees about how this order is explicitly applied~\cite{hector2010analysis,rasmussen2024_permutation_asca}, where the most simple approach is to assume an explicit order in the terms, e.g., $A$-$B$-$C(A)$-$AB$. Thus, $SS_A$ = $SS_T-E(A)$, $SS_B$ = $E(A)-E(A,B)$, $SS_{C(A)}$ = $E(A,B)-E(A,B,C(A))$ and $SS_{AB}$ = $E(A,B,C(A))-E(A,B,C(A),AB)$. In Type II SS, the SS of a term is adjusted for all other effects in the model of lower or equal order. An alternative definition (in the Matlab routine anovan) is to compute the reduction in residual SS obtained by adding that term to a model consisting of all other terms that do not contain the term in question. Both definitions, while slightly different, do match for the example at hand: $SS_A$ = $E(B)-E(A,B)$, $SS_B$ = $E(A)-E(A,B)$, $SS_{C(A)}$ = $E(A,B,AB)-E(A,B,C(A),AB)$ and $SS_{AB}$ = $E(A,B,C(A))-E(A,B,C(A),AB)$. In Type III SS, the SS for any given effect is calculated after all other effects in the model (regardless of order or hierarchy) have been accounted for. This is often applied along with the sigma restrictions (also known as "sum-to-zero" constraints), so that the term which SS is being calculated is constrained to sum 0. Type III SS in the example follow: $SS_A$ = $E(B,C(A),AB)-E(A,B,C(A),AB)$, $SS_B$ = $E(A,C(A),AB)-E(A,B,C(A),AB)$, $SS_{C(A)}$ = $E(A,B,AB)-E(A,B,C(A),AB)$ and $SS_{AB}$ = $E(A,B,C(A))-E(A,B,C(A),AB)$. All methods have advantages and disadvantages. Type I SSs are the easiest to compute but they depend on the specific order selected, so that different orders may lead to significantly different SSs, and so of estimated effects sizes, which are very important for data interpretation. Yet, resulting SSs exactly sum to the total $SS_x$, which while a nice behavior it also masks the imbalance problem in the data. Type II SSs are specially suitable in situations with large main effects and less relevance in the higher order terms, and they do not need to sum to the total $SS_T$. Type III SSs give the same \textit{a-priori} relevance to all terms, and they do not need to sum to the total $SS_x$.}

\section{Maximum likelihood estimation}

\subsection{Maximum likelihood in Linear Mixed Models}

In the regression equation for the maximum likelihood method, the fixed and random factors are separated into different parts. 
\begin{equation}
\mathbf{y} = \mathbf{X} \mathbf{\Theta} + \mathbf{Z} \boldsymbol{\gamma} + \boldsymbol{\epsilon}
\label{eq:Mixed model1}
\end{equation}
Here $\mathbf{X}$ is the coding for the fixed factors and $\mathbf{Z}$ for the random factors with $\mathbf{\theta}$ and $\boldsymbol{\gamma}$ their coefficients. The random factors should be considered as factors putting structure in the residuals. Similar to the residuals $\boldsymbol{\epsilon}$, the estimates for the levels of the random factor are considered to be normally distributed with mean $0$ and covariance $\boldsymbol{\Phi}$.
\begin{eqnarray}
    \label{Eq:Mixed model2}
    \boldsymbol{\gamma} \sim N
    \begin{pmatrix}
            0 , \boldsymbol{\Phi} 
    \end{pmatrix} \\ \nonumber
    \boldsymbol{\epsilon}\sim N
    \begin{pmatrix}
            0 ,\mathbf{\Sigma} 
    \end{pmatrix}\\ \nonumber
\end{eqnarray}
The covariance matrices ${\mathbf{\Phi}}$ and ${\mathbf{\Sigma}}$ are usually estimated using REstricted Maximum Likelihood (REML) and the final covariance matrix $\mathbf{{V}} = \mathbf{Z}{\Phi}\mathbf{Z}^T +  {\mathbf{\Sigma}}$ can be used in a weighted regression model to estimate the regression coefficients for the fixed factors $\mathbf{\Theta}$. REML is a statistical method for fitting Linear Mixed Models (LMMs) that corrects for bias in variance component estimates~\cite{montgomery2020design}.
\begin{equation}
\hat{\mathbf{\Theta}} = (\mathbf{X}^T \mathbf{\hat{V}}^{-1} \mathbf{X})^{-1} \mathbf{X}^T\mathbf{\hat{V}}^{-1} \mathbf{y}
\label{eq:Mixed model3}
\end{equation}
The REML estimates are expected to be more robust in the case of unbalanced data~\cite{martin_limmpca_2020} but they are making the additional assumption that all random factors follow approximately a normal distribution .

\subsubsection{Repeated Measures}

Lets exemplify a repeated measures model. The data from a randomized controlled trial (RCT) with a treatment group and a control group and 20 subjects per group that are measured at baseline and after 2 and 5 weeks can be modeled with a mixed model. The treatment groups are the fixed factor as the main interest is in the effect of the treatment. Time is a second fixed factor as one is interested in how fast the treatment effect becomes visible. The subject factor is a random factor. In this RCT we are not interested in each subject specifically, but in the population effect. The RCT assumes a random allocation of subjects in the groups, thus at baseline no difference between control and treatment groups can be expected. The model for such an experiment is:
\begin{equation}
y_{igt} = (\beta_0 + \gamma_{i}) + \beta_1 T_{1t} + \beta_2 T_{2t} + \beta_3(T_{1k}*G_g) + \beta_4(T_{2k}*G_g) + \epsilon_{igt}
\label{eq:Repeated_measures1}
\end{equation}
In this model $y_{igt}$ is the measurement for subject $i$, belonging to group $g$ measured at time point $t$. There is an overall intercept $\beta_0$ and a subject specific intercept $\gamma_{i}$, which is the random intercept. For time points 2 and 5 weeks, there is a general effect, which is independent of the group, $\beta_1 T_{1t}$ and $\beta_2 T_{2t}$, and there are group specific effects (or interactions) $\beta_3(T_{1k}*G_g)$ and $\beta_4(T_{2k}*G_g)$. Note that there is no general group effect. This is because at the first time point (the baseline) there is no difference between the groups due to the definition of the RCT. Finally, the residuals $\epsilon_{igt} \sim N(0,\sigma_e^2)$.

In a repeated measures model with a random intercept, the covariance matrix typically assumed is known as the Compound Symmetry (CS) covariance structure. This structure implies:
{e}qual variances {(t}he variances of the repeated measures $\sigma_u^2$ are equal across all time points{) and e}qual covariances {(t}he covariances between any two time points are equal{)}.
The compound symmetry covariance matrix for a subject $i$ then becomes:

\begin{eqnarray}
\mathbf{V}_i =  
 \left[
 \begin{array}{ccc}
 \sigma_e^2 +\sigma_u^2 & \sigma_u^2 & \sigma_u^2 \\
 \sigma_u^2 & \sigma_e^2 +\sigma_u^2  & \sigma_u^2 \\
 \sigma_u^2 & \sigma_u^2 & \sigma_e^2 +\sigma_u^2  \\
  \end{array}
 \right]
\label{eq:Repeated_measures2}
\end{eqnarray}

and for the whole $\mathbf{y}$ vector, $\mathbf{V}$ is a matrix with blocks $\mathbf{V}_i$ on its diagonal.
For the regression model notation of equation~\ref{eq:Mixed model1}: the coding for $\mathbf{X}$ and $\mathbf{Z}$ are provided in tables \ref{RepeatedMeasures_Ref Coding matrix} and \ref{RepeatedMeasures_Random Coding matrix}. If subject $i$ in a control subject, the upper part of table \ref{RepeatedMeasures_Ref Coding matrix} applies while for a treated subject the lower part applies.

\begin{table}[h]
\caption{Reference Coding matrix fixed factors $\mathbf{X}_i$}
    \label{RepeatedMeasures_Ref Coding matrix}
    \centering 
    \begin{tabular}{ccccccc}
        \toprule
        \textbf{Group} & \textbf{Time} & $\mu$ & $T_1$ & $T_2$ & $T_1 G$ & $T_2 G$ \\
        \midrule
        Control & $T_0$ & 1 &  0 &  0 & 0 &  0  \\
        Control & $T_1$ & 1 &  1 &  0 & 0 &  0  \\
        Control & $T_2$ & 1 &  0 &  1 & 0 &  0  \\ 
        ...     & ...     &...& ...&... &...&...  \\
        Treated & $T_0$ & 1 &  0 &  0 & 0 &  0  \\
        Treated & $T_1$ & 1 &  1 &  0 & 1 &  0  \\
        Treated & $T_2$ & 1 &  0 &  1 & 0 &  1  \\
        \bottomrule
    \end{tabular}
 \end{table}

\begin{table}[h]
\caption{Coding matrix random factors $\mathbf{Z}$}
    \label{RepeatedMeasures_Random Coding matrix}
    \centering 
    \begin{tabular}{ccccc}
        \toprule
        \textbf{Subject} & \textbf{Time} & $\gamma_1$ & $\gamma_i$ &  $\gamma_I$\\
        \midrule
        $1$ & $T_0$ & 1  &  0 &  0   \\
        $1$ & $T_1$ & 1  &  0 &  0   \\
        $1$ & $T_2$ & 1  &  0 &  0   \\
        ..  & ..    & .. & .. & ..   \\
        $i$ & $T_0$ & 0  &  1 &  0   \\
        $i$ & $T_1$ & 0  &  1 &  0   \\
        $i$ & $T_2$ & 0  &  1 &  0   \\
        ..  & ..    & .. & .. & ..   \\
        $I$ & $T_0$ & 0  &  0 &  1   \\
        $I$ & $T_1$ & 0  &  0 &  1   \\
        $I$ & $T_2$ & 0  &  0 &  1   \\
        \bottomrule
    \end{tabular}
 \end{table}
 
\newpage
\afterpage{\clearpage}
\bibliography{Bibliography}
\bibliographystyle{ieeetr}

\end{document}